\documentclass[useAMS,usenatbib]{mn2e}
\usepackage{times}
\usepackage{epsfig}
\usepackage{graphicx}
\usepackage{subfigure}
\usepackage{latexsym}
\usepackage{pifont}
\input epsf

\newcommand{\xmm}{{\it XMM~\/}}
\newcommand{\xmmn}{{\it XMM-Newton~\/}}
\newcommand{\asca}{{\it ASCA~\/}}

\newcommand{\bepposax}{{\it BeppoSAX~\/}}
\newcommand{\chandra}{{\it Chandra~\/}}

\newcommand{\rosat}{{\it ROSAT~\/}}
\newcommand{\einstein}{{\it Einstein~\/}}
\newcommand{\aap}{{\it A\&A~\/}}

\newcommand{\aj}{{\it AJ~\/}}
\newcommand{\apj}{{\it ApJ~\/}}
 
\newcommand{\apjs}{{\it ApJS~\/}}
\newcommand{\araa}{{\it ARA\&A~\/}}
\newcommand{\mnras}{{\it MNRAS~\/}}

\def\Msun{\hbox{$\rm ~M_{\odot}$}}
\def\ergcms{{\rm ~erg~cm^{-2}~s^{-1}}}

\def\ergsec{{\rm ~erg~s^{-1}}}
\def\cm2{{\rm ~cm^{-2}}}

\def\H0{{\rm ~km~s^{-1}~Mpc^{-1}}}

\def\eg{{\it e.g.~\/}}

\def\ie{{\it i.e.~\/}}

\def\la{\mathrel{\hbox{\rlap{\hbox{\lower4pt\hbox{$\sim$}}}{\raise2pt\hbox{$<$}}}}}
\def\ga{\mathrel{\hbox{\rlap{\hbox{\lower4pt\hbox{$\sim$}}}{\raise2pt\hbox{$>$}}}}}
\def\d25{D$_{25}$}
\def\nh{{$N_{H}$}}
\def\Ha{{H$\alpha$}}

\def\lx{L$_{\rm X}$}

\def\deg{\hbox{$^\circ$~\/}}
\def\arcm{\hbox{$^\prime$~\/}}
\def\arcs{\hbox{$^{\prime\prime}$~\/}}

\def\eps@scaling{1.0}%
\newcommand\plottwo[2]{%
  \centering
  \leavevmode
  \columnwidth=.45\columnwidth
  \includegraphics[width={\eps@scaling\columnwidth}]{#1}%
  \hfil
  \includegraphics[width={\eps@scaling\columnwidth}]{#2}%
}%

\title[X-ray emission from spiral galaxies] 
{X-ray emission from the extended disks of spiral galaxies}
\author[R.A.Owen  \& R.S. Warwick]
	{R.A.\ Owen, R.S.\ Warwick \\
X-ray \& Observational Astronomy Group, Dept. of Physics \& Astronomy, 
University of Leicester, University Road, Leicester LE1 7RH, U.K.\\}

\pagerange{\pageref{firstpage}--\pageref{lastpage}}

\pubyear{2009}

\begin{document}

\maketitle

\label{firstpage}


\begin{abstract}


We present a study of the X-ray properties of a sample of six nearby
late-type spiral galaxies based on \xmmn observations. Since our primary
focus is on the linkage between X-ray emission and star formation in extended,
extranuclear galactic disks, we have selected galaxies with near face-on
aspect and sufficient angular extent so as to be readily amenable to investigation
with the moderate spatial resolution afforded by {\it XMM-Newton}.  After excluding
regions in each galaxy dominated by bright point sources, we study both
the morphology and spectral properties of the residual X-ray emission,
comprised of both diffuse emission and the integrated signal
of the fainter discrete source populations. The soft X-ray morphology generally
traces the inner spiral arms and shows a strong correlation with
the distribution of UV light, indicative of a close connection between the
X-ray emission and recent star formation. The soft (0.3--2 keV) X-ray luminosity
to star formation rate (SFR) ratio varies from
$1-5 \times10^{39}\ergsec$ (\Msun~yr$^{-1}$)$^{-1}$, with an indication
that the lower range of this ratio relates to regions of lower SFR density.
The X-ray spectra are well matched by a two-temperature thermal model
with derived temperatures of typically $\sim 0.2$ keV and $\sim 0.65$ keV, in line 
with published results for other normal and star-forming galaxies. The hot component 
contributes a higher fraction of the soft luminosity in the galaxies with 
highest X-ray/SFR ratio, suggesting a link between plasma temperature and 
X-ray production efficiency. The physical properties of the gas present in the galactic
disks are consistent with a clumpy thin-disk distribution, presumably composed 
of diffuse structures such as superbubbles together with the integrated emission 
of unresolved discrete sources including young supernova remnants.

\end{abstract}



\section{Introduction}
\label{sec:intro}

The first detailed studies of the X-ray emission from nearby late-type galaxies
were carried out with the \einstein observatory and revealed both individual luminous 
X-ray sources, as well as an extended, often complex, underlying structure
(\citealt{fabbiano89}). In a number of near edge-on star-forming systems. such
as M82 (\citealt{watson84}), evidence was also found for hot outflowing winds.
The superior spatial resolution and enhanced soft X-ray response of \rosat greatly 
extended this work, leading to the detection of substantial numbers of discrete
sources as well as establishing a much clearer picture of the diffuse emission
components (\eg \citealt{read97}). More recently the advent of \xmmn and {\it Chandra} 
has extended our knowledge of the discrete X-ray source populations and the ultrahot ISM 
in nearby galaxies to unprecedented luminosity and surface brightness thresholds
(\eg \citealt{strickland04a}; \citealt{colbert04}; \citealt{pietsch04}; 
\citealt{fabbiano06}; \citealt{yang07};  \citealt{bogdan08}; \citealt{bauer08}).
 
The diffuse X-ray emission associated with the disks of spiral galaxies 
is believed to arise in gas heated by shocks generated as a result of supernova
explosions and stellar wind interactions  (\citealt{chevalier85};  \citealt{strickland00}; 
\citealt{suchkov94}; \citealt{grimes05}). This is supported by analysis of the 
X-ray spectra 
of individual shell-type supernova remnants (SNRs) of reasonable age, which are often
characterised by plasma temperatures in the range 0.2-0.8 keV, similar to the values
obtained when studying larger-scale structures in galactic disks. 
\citet{strickland04a} showed that the luminosity of the diffuse X-ray 
emission observed in star-forming galaxies is proportional to the rate of mechanical 
energy injection into the ISM from young stars. \citet{mas08} further found that the 
ratio of the soft X-ray luminosity to far infrared luminosity, taking the latter as a 
proxy for the star-formation rate (SFR), is strongly dependent on the efficiency 
with which 
mechanical energy is converted into soft X-rays and also the evolutionary status of the 
star formation episode. This all argues for a close connection between soft X-ray emission 
and star formation in late-type galaxies.


\begin{table*}
\caption{Properties of the sample of galaxies.}
\centering
\begin{tabular}{lcccccc}
\hline
Galaxy    & Hubble Type~$^{a}$   &  Distance~$^{b}$ & \d25~$^{a}$ & Inclination~$^{b}$  & 
Foreground \nh~$^{c}$  & Nucleus \\
& & (Mpc) & ($\arcm$) &   ($^{\circ}$) & ($10^{20}\cm2$) & \\
\hline
NGC300  & SA(s)d     & 2.0    & 21.8  & 30  & 3.6 & - \\
M74     & SA(s)c     & 11.0   & 10.5  & 7   & 4.8 & - \\
NGC3184 & SAB(rs)cd    & 11.6   & 7.2   & $<24$  & 1.0 & - \\
M51     & SA(s)bc    & 8.4    & 7.3   & 20  & 1.8 & Sy 2\\
M83     & SAB(s)c     & 4.5    & 12.9  & 24  & 3.9 & Starburst \\
M101    & SAB(rs)cd    & 7.2    & 26.9  & 18 & 1.1 & - \\

\hline    
\end{tabular}
\\
\raggedright
$^{a}$ - Hubble type and the major-axis diameter (\d25) from the RC3 
catalogue (\citealt{devaucouleurs91}). \\
$^{b}$ - See \S\ref{sec:morphology} for references. \\
$^{c}$ - \nh~values based on \citet{dickey90}.

\label{table:gal:details}
\end{table*}


\begin{table*}
\caption{Details of the \xmmn observations of the five new galaxies in our sample.}
 \centering
  \begin{tabular}{lcccccccccc}
\hline
Galaxy    & Observation ID & Start Date  & Filter~$^{a}$   & & \multicolumn{2}{c}{Target co-ordinates~$^{b}$}  & & \multicolumn{2}{c}{Useful exposure (ks)} \\
      &                & (yyyy-mm-dd) & pn/MOS1/MOS2    & & RA (J2000)       & Dec (J2000)             & & pn            & MOS 1+2
    \\
\hline
NGC300  &  0112800201  &  2000-12-26  &  M/M/M   & & $00^h54^m55.0^s$ & $-37\deg41\arcm00\arcs$ & & 29.4  &  65.1                     \\
   & 0112800101     & 2001-01-01   & M/M/M  & &   &  & & 40.4  & 88.4   \\
  &  0305860401  &  2005-05-22  &  M/M/M  & &   &   & &  27.2  &  64.9  \\
  &  0305860301  &  2005-11-25  &  M/M/M  & &   &   & &  30.4  &  68.2  \\
M74  &  0154350101  &  2002-02-02  &  T/T/T   & & $01^h36^m23.9^s$ & $+15\deg45\arcm13\arcs$ & & 23.3   &  49.9                   \\
 & 0154350201     & 2003-01-07   & T/T/T   & &  &   & & 23.5  & 50.2     \\
NGC3184 & 0028740301     & 2001-11-02   & T/T/T   & & $10^h18^m40.0^s$ & $+41\deg25\arcm11\arcs$ & & 19.9          & 49.9                     \\
M51     & 0112840201  &  2003-01-15   &  T/T/T     & & $13^h29^m51.9^s$ & $+47\deg10\arcm32\arcs$ & & 18.8          & 40.6                     \\
  &  0212480801  &  2005-07-01   &  M/M/M   & &   &   & &  20.4  &  56.7  \\
  & 0303420101     & 2006-05-20   & T/T/T  & &   &   & &  32.3   & 77.3  \\         
  &  0303420201  &  2006-05-24   &  T/T/T   & &   &   & &  14.1  &  43.0  \\
M83     & 0110910201     & 2003-02-18   & T/M/M   & & $13^h37^m00.4^s$ & $-29\deg52\arcm04\arcs$ & & 18.3          & 38.2                     \\
\hline
\end{tabular}
\\
\raggedright
$^{a}$ - T = thin filter, M = medium filter. \\
$^{b}$ - Assumed position of the galactic nucleus.\\
\label{table:obs}
\end{table*}


A recent study of a sample of 12 nearby spiral galaxies using \chandra 
(\citealt{tyler04}) has shown strong correlation between soft diffuse X-ray emission and 
sites of recent star formation in spiral arms traced by mid-infrared and \Ha~emission. 
A similar investigation has not been completed using {\it XMM-Newton}, but  
a number of studies of individual spiral galaxies have been 
reported (\citealt{pietsch01};
\citealt{takahashi04}; \citealt{trudolyubov05}; \citealt{warwick07}), 
which again demonstrate the strong linkage between tracers of star formation and 
diffuse X-ray emission. Furthermore, a dual-component thermal model, with 
characteristic temperatures of $\sim 0.2$ keV and 0.6--0.7 keV, very often 
provides a good description of the spectrum of the diffuse X-ray emission. 
Studies of the point source X-ray luminosity function (XLF) in spiral galaxies
(\citealt{tennant01}; \citealt{soria03}; \citealt{colbert04}) have shown that the 
brightest 
point sources contain the bulk of the integrated point source luminosity, suggesting
that, provided a sufficient number of the brightest point sources are excluded, 
it should in principle be possible to probe the underlying structure of the galaxy to 
relatively low levels of surface brightness.

In this paper, we focus on the spectral and morphological properties of the 
diffuse X-ray emission emanating from the extended disks of late-type normal 
galaxies, as deduced from \xmmn observations. In \S\ref{sec:sample}, we 
identify a sample of nearby galaxies with close to face-on disks of sufficient 
angular extent so as to be readily amenable for study with {\it XMM-Newton}. 
We then outline the data reduction methods 
employed (\S\ref{sec:obs}).  Using image analysis, we sub-divide the total X-ray 
luminosity of each galaxy into the contribution of spatially resolved bright sources
and a {\it residual} component comprising the integrated emission of unresolved 
point sources and the diffuse galactic emission (\S\ref{sec:lx}). On the basis of
published source luminosity functions, we estimate the likely contribution of truly 
diffuse emission to this {\it residual} signal.  We go on to investigate spectral 
properties of both 
the bright source and the unresolved (residual) components (\S\ref{sec:spectrum}). 
We also consider the morphology of the residual emission in the context of the 
star formation evident in the galactic disks (\S\ref{sec:star}). Finally we discuss 
the implication of our results (\S\ref{sec:disc})   and briefly summarise our conclusions 
(\S\ref{sec:conc}).

\section{Galaxy Sample and Observations}
\label{sec:sample}

In this paper we investigate the large-scale spatial and spectral properties 
of the X-ray emission emanating from the extended disks of six nearby late-type
spiral galaxies. The galaxies in question are NGC300, M74, NGC3814, M51, M83 and 
M101, brief details of which are given in Table~\ref{table:gal:details}. In a previous
paper (\citealt{warwick07}; hereafter W07), we reported the results of an \xmmn study of
one of these galaxies, namely M101, and here we apply a similar methodology 
to extend the sample, thereby allowing the intercomparison of the galaxy
X-ray characteristics based on \xmmn observations.

When dealing with X-ray spectral imaging data of only moderate spatial resolution 
(as afforded by {\it XMM-Newton}), the presence of a number of 
very bright point sources has the potential to obscure the properties of the 
underlying galaxy.
Since our goal is to study the latter rather 
than the former, we have limited our galaxy sample to nearby face-on systems
of Hubble type Sbc-Sd with major-axis \d25 extent greater than 7\arcm, thereby
ameliorating the source confusion problem.
(We have excluded M33 from our study since this galaxy has a \d25 diameter of 
70\arcm and thus extends well beyond the EPIC field of view; M33 is the subject of 
a specific \xmmn programme utilising multiple pointings (\citealt{pietsch04};
\citealt{misanovic06}). 

Details of the \xmmn  EPIC observations of the five new additions
to our galaxy sample are summarised in Table~\ref{table:obs}. 


\section{Data Reduction}
\label{sec:obs}
 
The data reduction was based on SAS v8.0.
The datasets were initially screened for periods of high background by
accumulating full-field 10--15\,keV light curves. MOS data were 
excluded during periods when the 10-15 keV count rate averaged over 100 s 
bins exceeded 0.2 $\rm ct~s^{-1}$, and the pn data were 
similarly excised 
when the pn count rate exceeded 2 $\rm ct~s^{-1}$.   
As summarised in Table~\ref{table:obs}, the exposure times after event 
filtering within individual observations ranged from 14 to 40 ks in the
pn camera (with typically slightly greater exposure in the MOS cameras).  

In W07, we discussed how the use of an
appropriate spatial mask greatly reduced the ``contamination'' arising
from the presence of relatively small numbers of luminous point sources
in the disk of M101. Here we employ a refinement of this masking technique 
in our analysis of the \xmmn observations of five further target galaxies. 

\subsection{Image Construction}
\label{sec:images}

The starting point of our analysis, for each galaxy and for each individual 
observation, was the creation of pn, MOS 1 and MOS 2 images and associated 
exposure maps in three energy bands: 
soft (0.3--1\,keV),  medium  (1--2\,keV)  and   hard  (2--6\,keV). In all cases,
the pixel size was set at $4'' \times 4''$. 
For both the image analysis and the subsequent spectral analysis, we
utilised single and double pixel events in the pn camera (pattern 0--4) and
single to quadruple events (pattern 0--12) in the MOS cameras. 
The resulting images were then flat-fielded by 
subtraction of a constant background particle rate (estimated from count rates 
recorded in the corners of the CCD detectors not exposed to the sky) and division 
by the relevant exposure map. At the same time, spurious data from bad pixels, 
hot columns and chip gaps were excluded. Where there was more than one observation, 
the images pertaining to a given instrument and band were at this point combined
into a single mosaiced image. For each band, the resultant flat-fielded MOS 1 and 
MOS 2 images were combined to give a ``MOS'' image and finally a ``pn+MOS'' image 
was also produced by summing the individual pn and mos images (with appropriate 
scaling in regions affected by chip gaps, where there was incomplete coverage 
in one of the instrument channels).  

\begin{figure*}
\centering
\rotatebox{270}{\scalebox{0.45}{\includegraphics{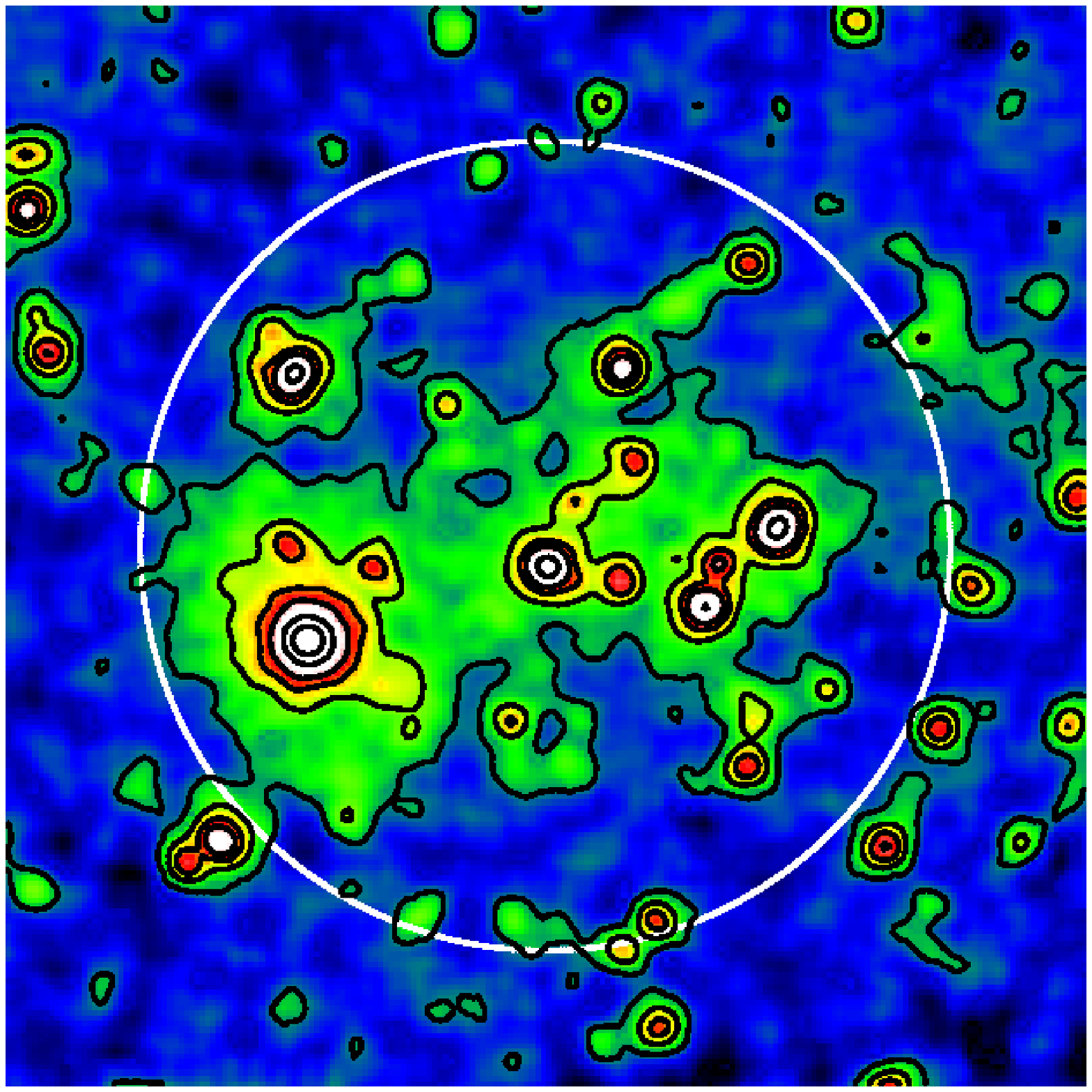}}}
\rotatebox{270}{\scalebox{0.45}{\includegraphics{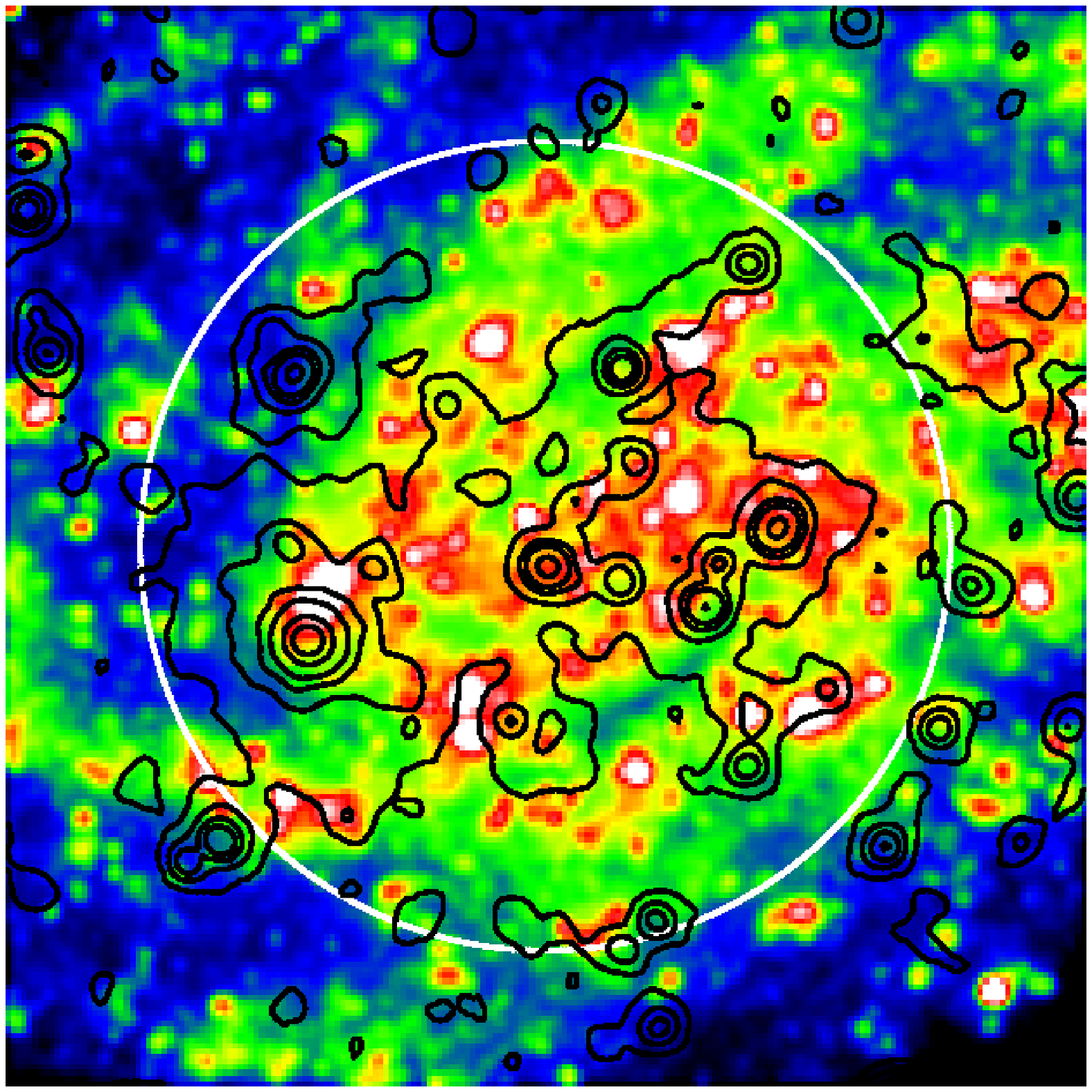}}}
\rotatebox{270}{\scalebox{0.45}{\includegraphics{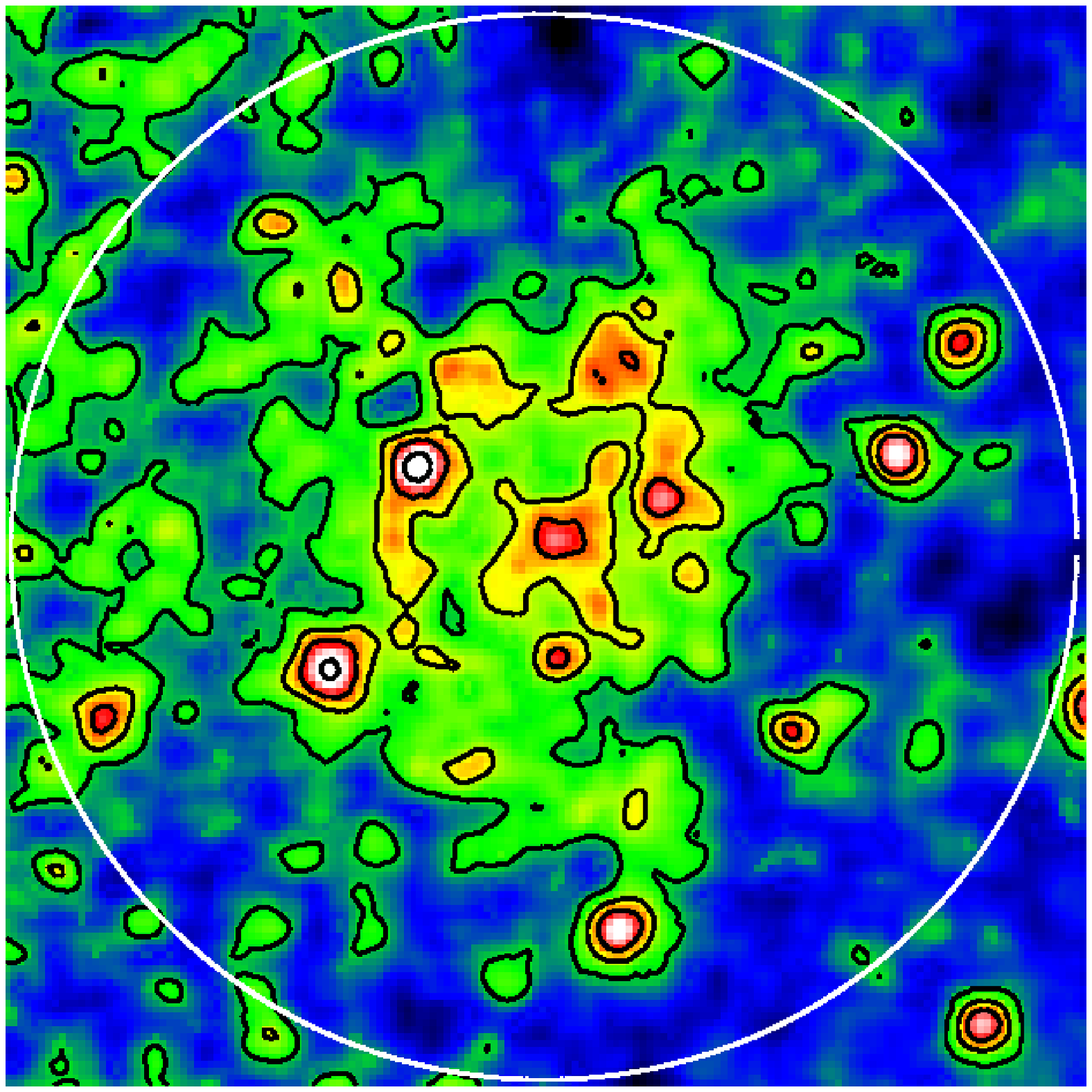}}}
\rotatebox{270}{\scalebox{0.45}{\includegraphics{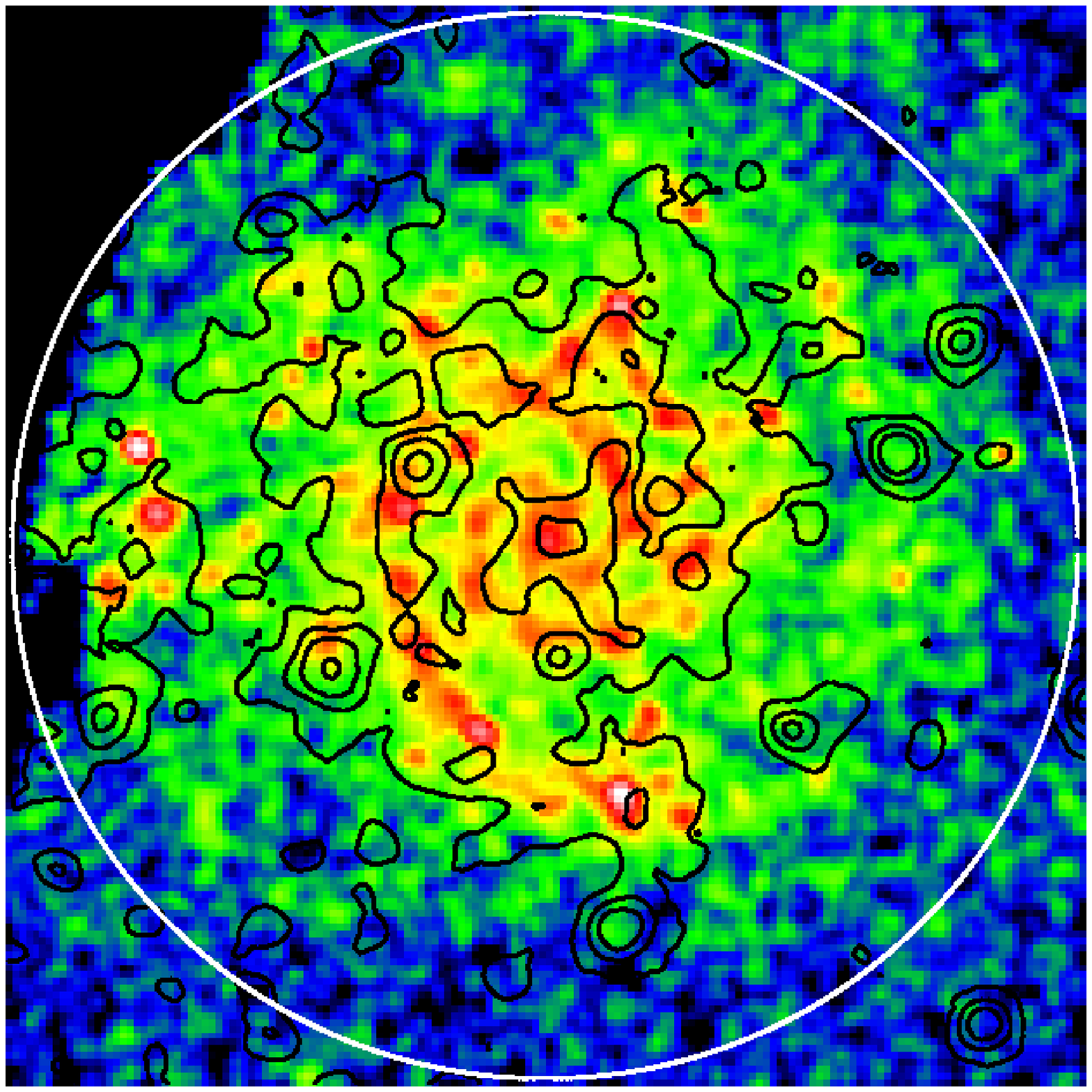}}}
\rotatebox{270}{\scalebox{0.45}{\includegraphics{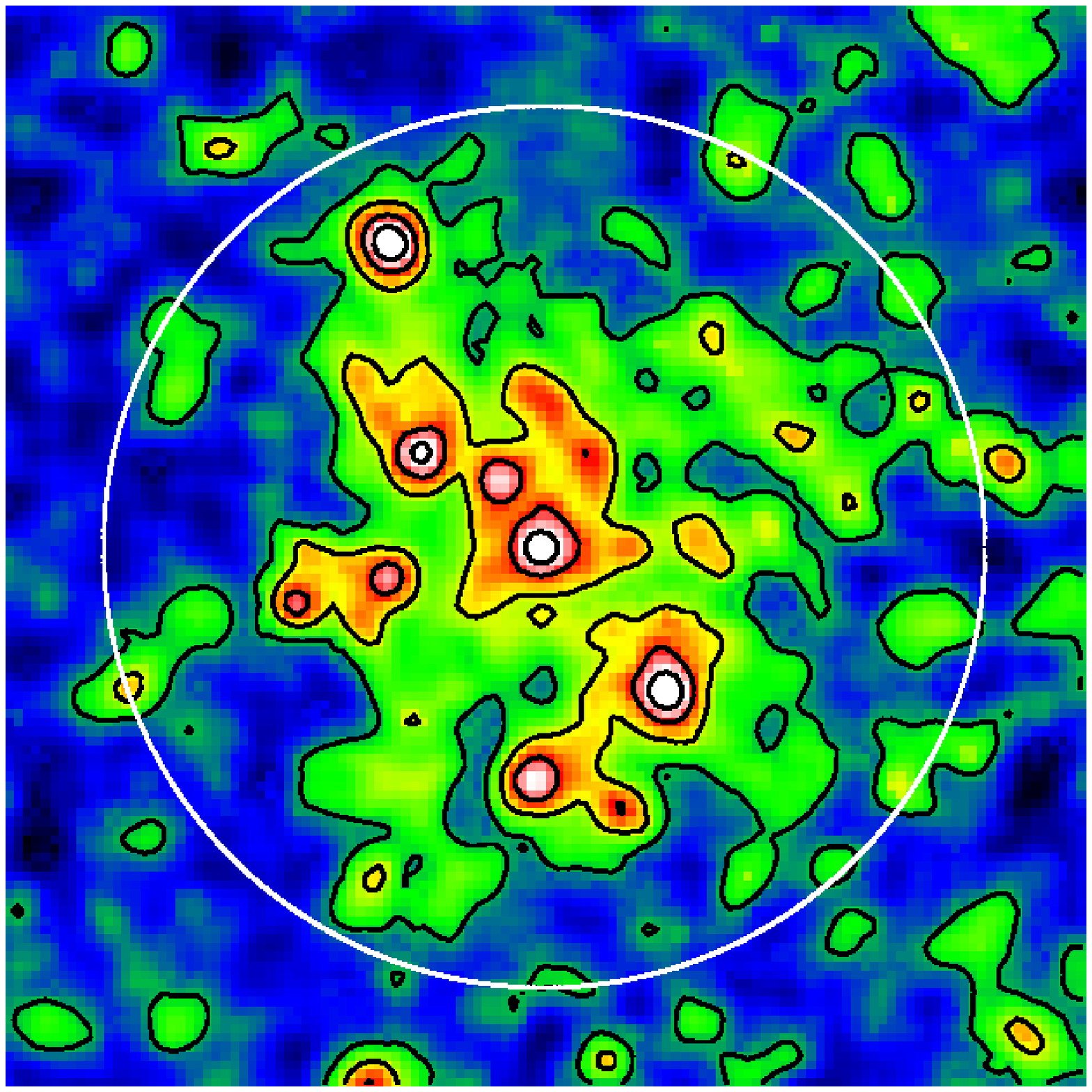}}}
\rotatebox{270}{\scalebox{0.45}{\includegraphics{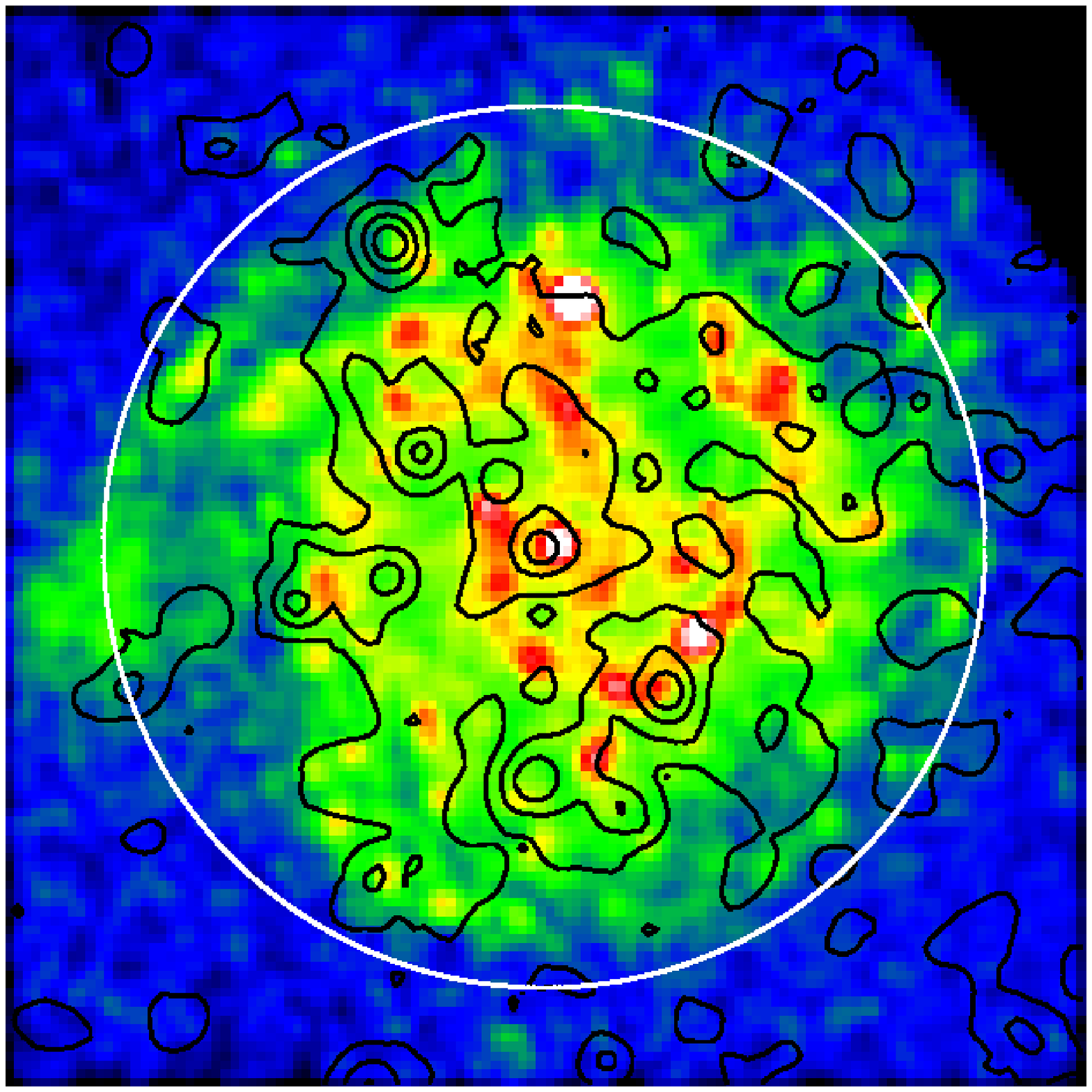}}}
\caption{Comparison of the soft X-ray and far-ultraviolet images
of three galaxies. {\it Left-hand panels:}  Adaptively smoothed versions of
the \xmmn pn+MOS images in the soft (0.3-1 keV) band. 
The amplitude scaling in all cases is logarithmic with the contour levels 
representing factor two steps in the soft X-ray surface brightness. 
{\it Right-hand panels:}  
The {\sl GALEX} FUV ($\lambda_{eff}=2267$ \AA) image on the same spatial 
scale as the X-ray data (except for NGC 3184 where the UV data are 
from the \xmmn Optical Monitor in the UVW1 ($\approx2680$ \AA)
band.  The amplitude scaling is again logarithmic. The contours show the 
soft X-ray morphology for comparison purposes.
{\it Top panels:} NGC 300.  The circle has a diameter of $10^{\prime}$.  
{\it Middle panels:} M74. The circle has a diameter of $10.5^{\prime}$.  
{\it Bottom panels:} NGC3184. The circle has a diameter of $7.2^{\prime}$.  
}
\label{fig:xuv1}
\end{figure*}

\begin{figure*}
\centering
\rotatebox{270}{\scalebox{0.45}{\includegraphics{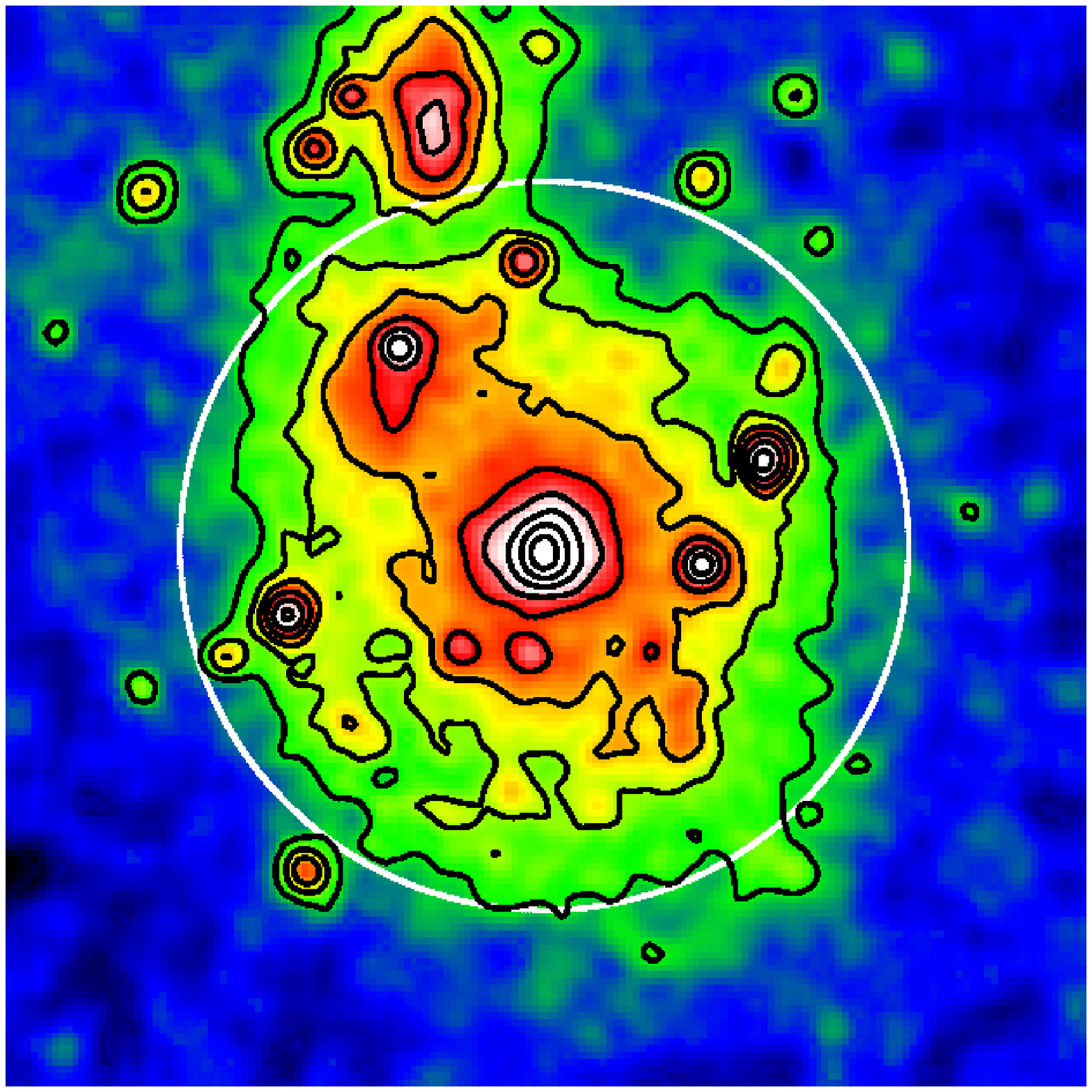}}}
\rotatebox{270}{\scalebox{0.45}{\includegraphics{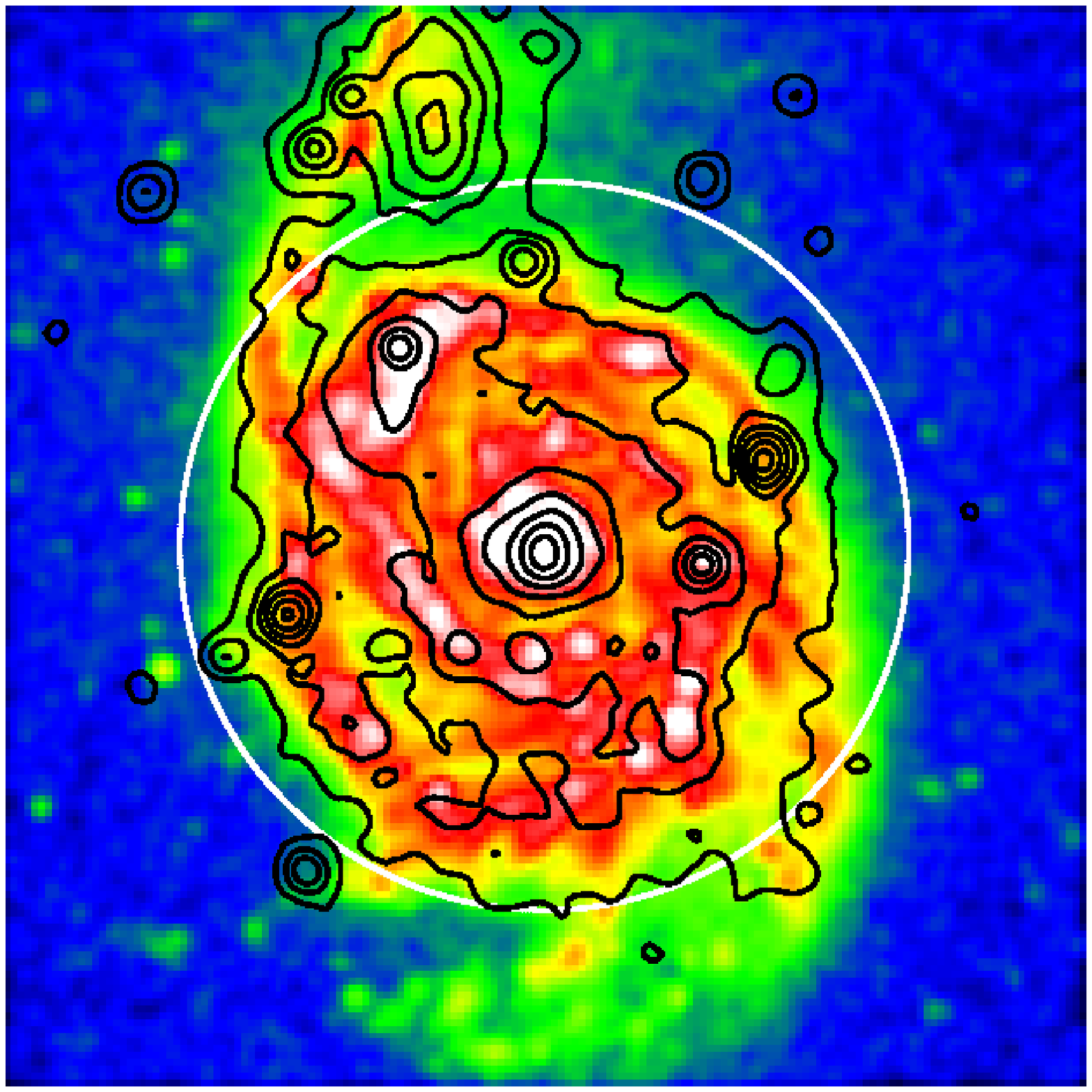}}}
\rotatebox{270}{\scalebox{0.45}{\includegraphics{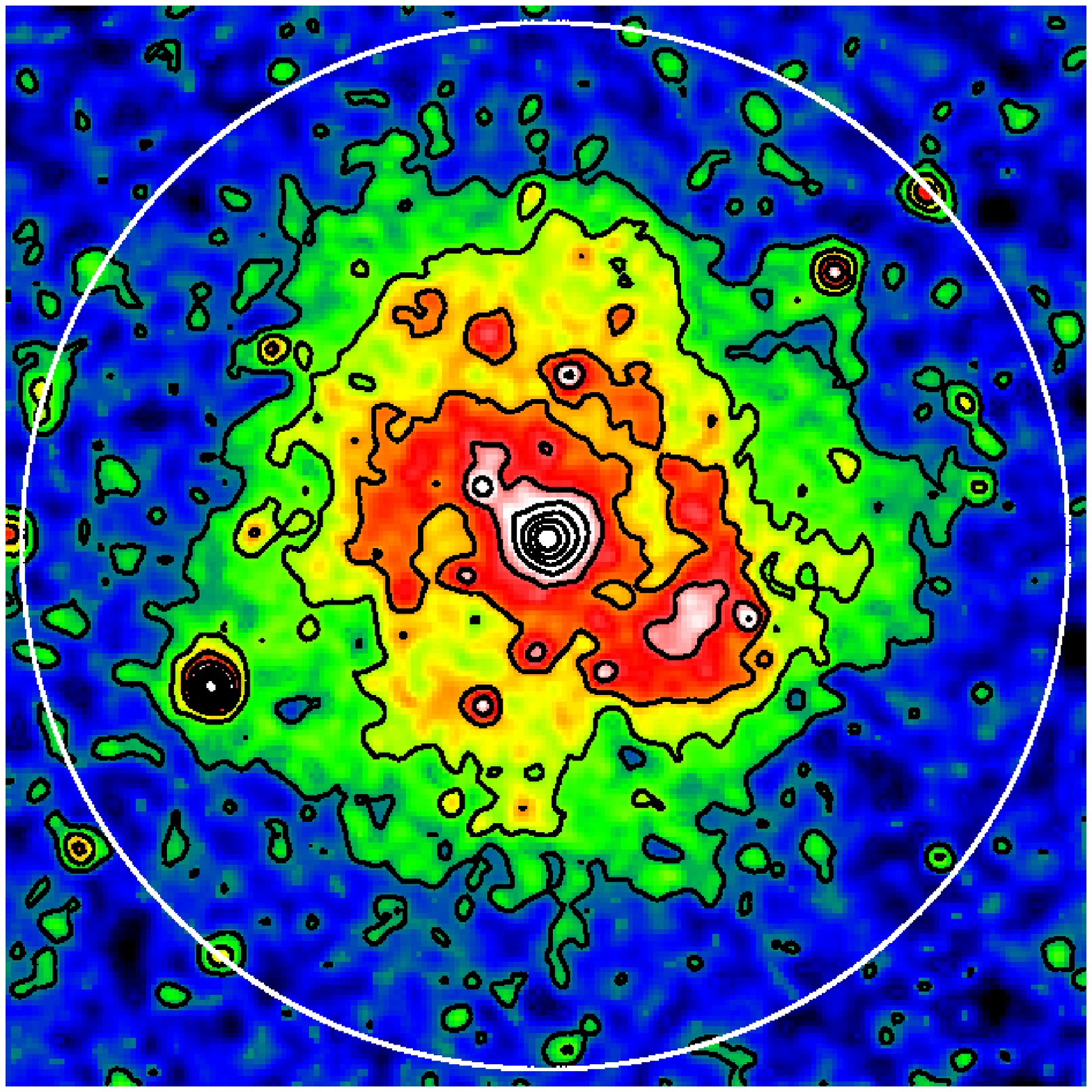}}}
\rotatebox{270}{\scalebox{0.45}{\includegraphics{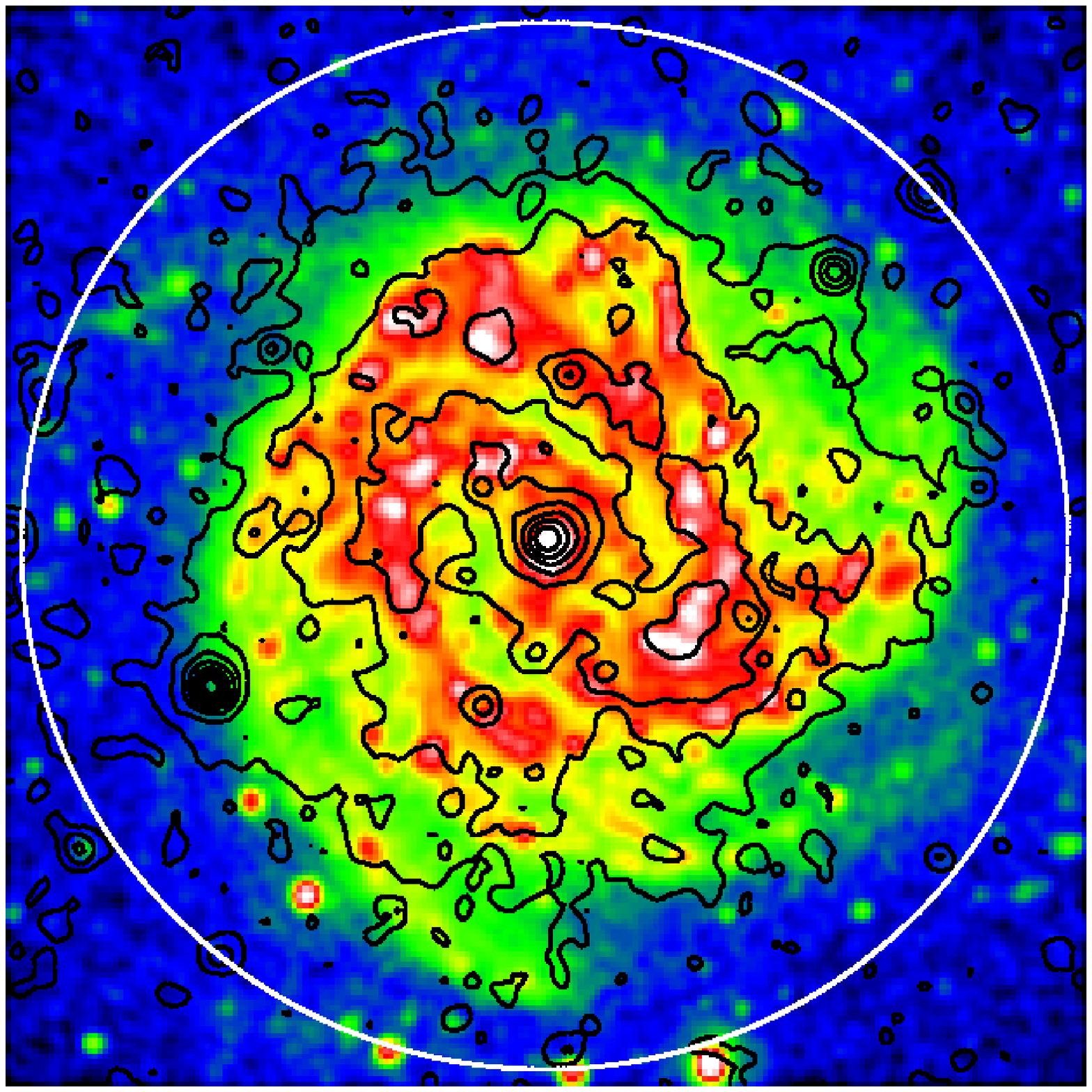}}}
\rotatebox{270}{\scalebox{0.45}{\includegraphics{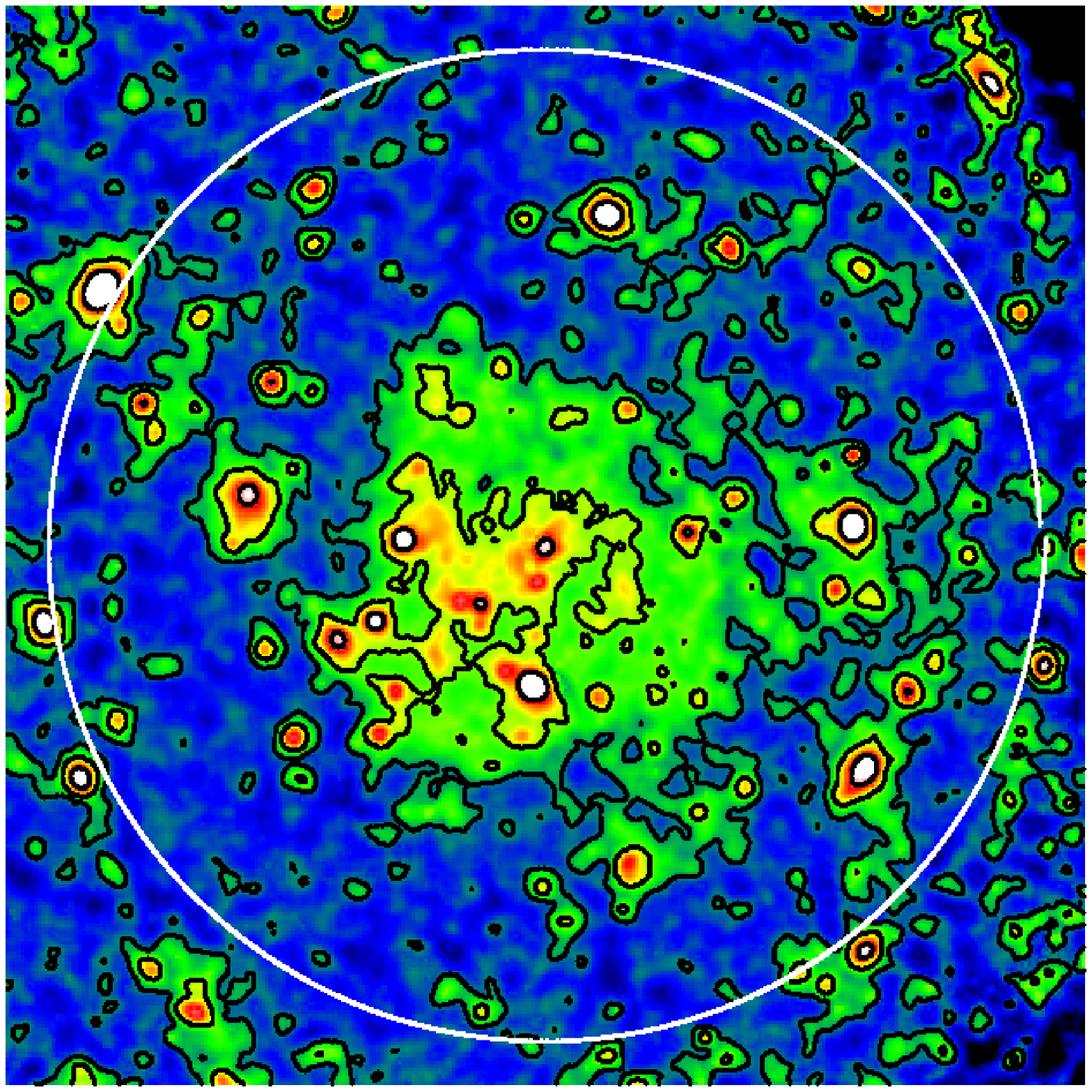}}}
\rotatebox{270}{\scalebox{0.45}{\includegraphics{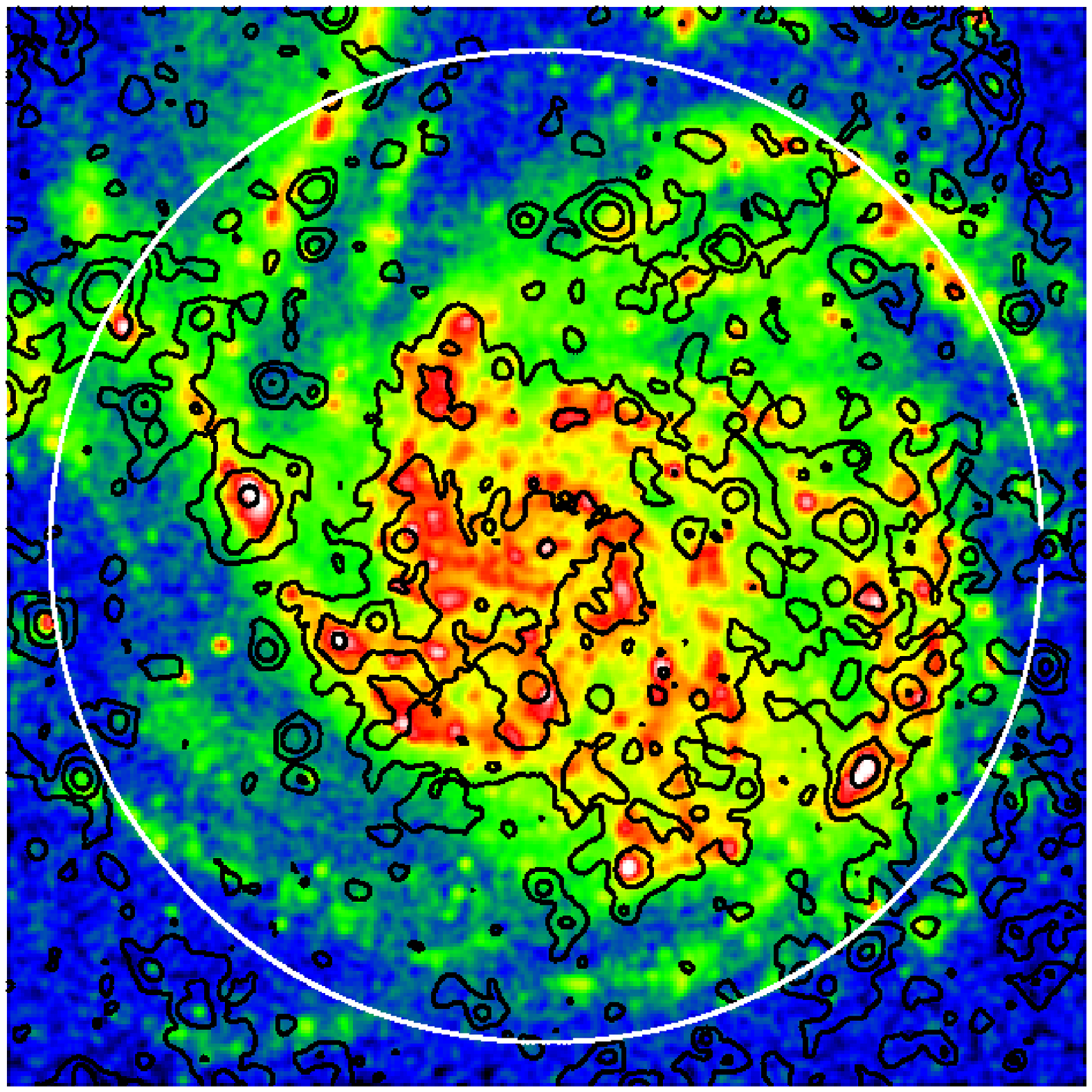}}}
\caption{As for Fig. \ref{fig:xuv1}.
{\it Top panels:} M51.  The circle has a diameter of $7.3^{\prime}$.   
{\it Middle panels:} M83.  The circle has a diameter of $12.9^{\prime}$.  
{\it Bottom panels:} M101.  The circle has a diameter of $20^{\prime}$.}
\label{fig:xuv2}
\end{figure*}

Figures \ref{fig:xuv1} \& \ref{fig:xuv2} 
show the soft-band images derived for each of the five new galaxies, together 
with the corresponding data for M101 from W07. In the case of M74, NGC3184, M51 and M83
our analysis pertains to circular regions centred on the galactic nucleus
of diameter equal to the major-axis \d25 dimension, whereas for NGC 300, given
the relatively low X-ray surface brightness, we restrict our attention to a 
central 10\arcm 
diameter region (0.46 of the \d25 extent). Similarly for M101 the X-ray analysis 
reported in W07 applies to a central 20\arcm diameter region (0.74 of the \d25
dimension). Hereafter we refer to these galaxy-centred circular regions as 
the ``X-ray extraction regions''.


\begin{figure*}
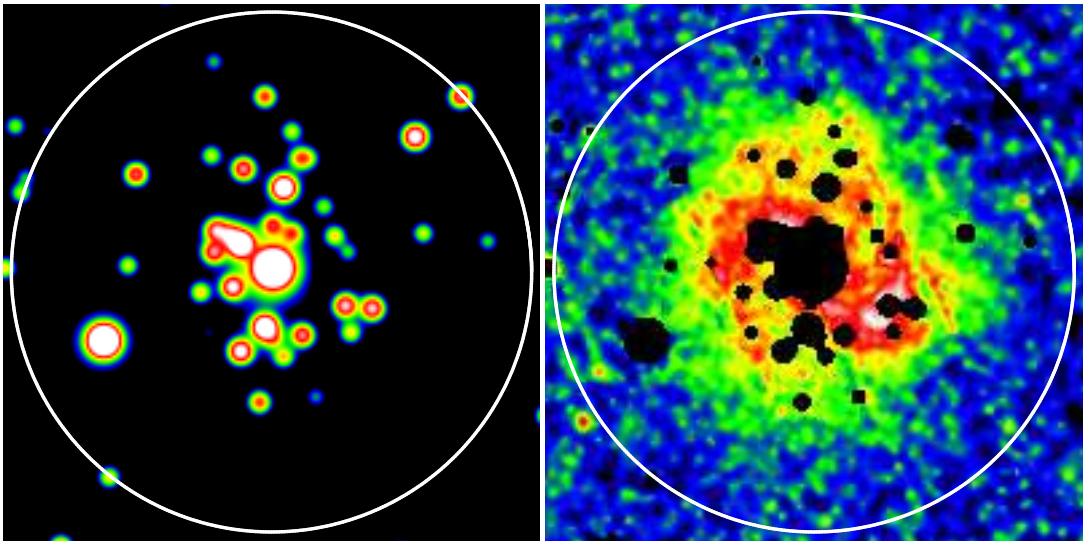

\centering
\rotatebox{270}{\scalebox{0.45}{\includegraphics{revimage1b.ps}}}
\rotatebox{270}{\scalebox{0.45}{\includegraphics{revimage1c.ps}}}
\caption{ {\it Left panel:}  The modelled ``bright source'' image constructed for M83
from a set of count-rate scaled PSF sub-images.  A source mask was produced by 
thresholding
this image at a surface brightness level such that $91\%$ of the source signal
was contained within the resulting mask.
{\it Right panel:} The soft-band image of the residual emission in M83 obtained
from the full image by subtracting the bright source model and then applying the 
spatial mask.
Much of the ``contamination'' due to bright sources is suppressed by this
process. In both panels the circle has a diameter of $12.9'$.}
\label{fig:im_big}
\end{figure*}


\subsection{Spatial masking of bright sources}
\label{sec:imagemask}

In order to produce a spatial mask to suppress the bright source contamination 
it was necessary to construct a catalogue of the bright sources associated with 
each galaxy.  The starting point was a preliminary source-list for each galaxy 
derived from the second \xmmn Serendipitous Source Catalogue (2XMM; \citealt{watson08}), 
encompassing those sources located within the galaxy extraction region (as defined above).  
Next we determined the net count-rate of each source in the broad-band pn+MOS image\footnote
{The pn+MOS1+MOS2 count-rate above the local background was determined within a cell of 
$16''$ radius centred on the source position. A correction factor was then applied to 
allow for the signal contained in the wings of the point spread function (PSF) extending 
beyond the nominal source cell.}.
Those sources for which the measured count rates exceeded a defined {\it count-rate} 
threshold (specific to each galaxy) were then included in the final ``master'' 
source-list for the galaxy.  In the case of both M83 and NGC 300 two additional sources, 
not included in the 2XMM catalogue, but clearly present in the images at count-rates 
above the threshold were added by hand. Similarly in the case of M51 one additional 
source was added. The number of sources included in each master 
source-list is given in Table \ref{table:gal_sources}, together with the threshold 
setting (quoted as the equivalent unabsorbed source luminosity in the 0.3--6\,keV band
on the basis of a source spectrum consisting of power-law continuum with 
$\Gamma=1.7$ absorbed by the Galactic foreground \nh).  At this stage we also
determined the count rates for each source in the soft, medium and hard bands
by applying the above process to the appropriate pn+MOS sub-band images.


\begin{table*}
\caption{Characteristics of the bright X-ray source population in each galaxy.}
\centering
\begin{tabular}{lccccccc}
\hline
Galaxy  &  X-ray Region$^{a}$ & Threshold~\lx$^{b}$ & Number in 
& \multicolumn{1}{c}{Number of high \lx~sources} \\
        & (\arcm) &  ($10^{37}\ergsec$) &  Source List
&  (\lx $> 5 \times 10^{38}\ergsec$) \\
\hline
NGC300   & 10.0  & 0.2 & 22  & 0  \\
M74      & 10.5   & 5.0 & 20  & 4 \\
NGC3184  & 7.2  & 6.0 & 18  & 4 \\
M51      & 7.3   & 4.0 & 26  & 6 \\
M83      & 12.9  & 2.0 & 40  &  2 \\
\hline    
\end{tabular}
\\
$^{a}$ - Diameter of the ``X-ray extraction region''. \\
$^{b}$ - Nominal \lx~threshold applied in the 0.3--6 keV band in defining sample. \\
\label{table:gal_sources}
\end{table*}


The next step involved the creation of the spatial mask. 
To this end, we produced a simulated image of the point sources present in the 
broad-band pn+MOS observation. The MOS- and pn-camera PSFs were modelled for each 
energy band as a set of sub-images representing the PSF out to $1'$ 
from the source position. A simulated sub-image of each source (in three bands 
and two instruments) was then produced by scaling the appropriate PSF image 
so as to match the actual image data (in terms of the 
count rate contained within the nominal source cell). 
The resultant set of sub-images were then co-added with 
suitable spatial offsetting so as to simulate the surface brightness distribution 
in the broad-band pn+MOS image deriving from bright point sources within the galaxy 
field\footnote{As a final iteration, count rates were extracted from the simulated images 
within a $16''$ cell centred on each source position and compared to those obtained 
from the actual data, with any differences attributable to the PSF spreading 
of signal from adjacent sources. Based on this analysis, the nominal source count 
rates were adjusted to compensate for this source confusion effect and a new 
version of the simulated images was produced.}. A spatial mask was then derived 
from this image by applying an appropriate surface brightness cut.

Figure \ref{fig:im_big} shows the simulated ``bright source image'' derived
using the above procedure for M83. In this example, the  bright source mask 
was obtained by applying a surface brightness cut to the image at a level of
0.07 pn+MOS1+MOS2 $\rm ct~ks^{-1}~pixel^{-1}$, leading to a ``spillover'' fraction,
({\it i.e.,} the fraction of the bright source signal not contained within the masked 
region) of 9\%.  Similar surface brightness cuts were applied 
to the other galaxies resulting in spill-over fractions in the range 
9-12\%. 

The derived spatial masks were used to separate the ``bright source region'' 
from the ``residual emission region'' in both the spatial and spectral analysis. 
In order to study the spatial distribution of the residual emission 
(see \S \ref{sec:star}), a simulated image representative of the soft band
data was subtracted from the corresponding pn+MOS image and the source 
mask imposed - see Figure \ref{fig:im_big}. This process serves to to suppress much 
of the ``contamination'' arising from the brightest sources in the galaxy, 
including the bulk of the spillover into the residual galaxy region.  Note that a slightly
different approach is necessary in the spectral analysis; here we take account 
of the bright source signal spilling into the residual emission region by including an 
appropriate fraction of the bright source spectral model in the spectral fit of the 
residual emission - see \S\ref{sec:spec:res}.


\begin{table*}
\caption{Contribution of point sources to the total X-ray luminosity.}
\centering
\begin{tabular}{lcccccc}
\hline

Galaxy &  Spillover/Area  & Component   & \multicolumn{4}{c}{\lx ($10^{39}\ergsec$)}
     \\
& Factors (\%) & & (0.3-1 keV) & (1-2 keV) & (2-6 keV) & (0.3-6 keV)   \\
\hline
NGC300 & 12/7 & Bright Sources & 0.19 & 0.07 & 0.25 & 0.51 \\
        &       & Unresolved Sources & [0.02] & [0.01] & [0.03] \\
       &      & Residual Galaxy & 0.05 &    - &    - & 0.05 \\
       &      & Total Measured  & 0.24 & 0.07 & 0.25 & 0.56 \\
\\
M74 &  11/24  & Bright Sources  & 1.4 & 1.3 & 3.0  & 5.7  \\
        &       & Unresolved Sources & [0.5] & [0.4] & [0.8] \\
    &         & Residual Galaxy & 1.4 & 0.7 & -    & 2.1  \\
    &         & Total Measured  & 2.8 & 2.0 & 3.0  & 7.8  \\
\\
NGC3184 & 9/22 & Bright Sources & 1.2 & 1.4 & 3.4 & 6.0 \\
        &       & Unresolved Sources & [0.6] & [0.5] & [0.9] \\
        &      & Residual Galaxy & 2.0 & 0.5 & -   & 2.5 \\
        &      & Total Measured  & 3.2 & 1.9 & 3.4 & 8.5 \\
\\
M51     & 12/24 & Bright Sources  & 6.8 & 3.8 & 6.5 & 17.1 \\
        &       & Unresolved Sources & [0.5] & [0.4] & [0.8] \\
        &       & Residual Galaxy & 6.3 & 0.9 & -   &  7.2 \\
        &       & Total Measured  & 13.1 & 4.7 & 6.5 & 22.3 \\
\\
M83     & 9/30  & Bright Sources  & 2.5 & 2.0 & 3.1 & 7.6 \\
        &       & Unresolved Sources & [0.5] & [0.4] & [0.9] \\
        &       & Residual Galaxy & 3.3 & 0.7 & -   & 4.0 \\
        &       & Total Measured    & 5.8 & 2.7 & 3.1 & 11.6 \\
\hline    
\end{tabular}
\label{table:gal_xray}
\end{table*}


\subsection{Spectral Extraction}
\label{sec:diff_extr}

The soft-band images shown in Fig.~\ref{fig:xuv1} \& \ref{fig:xuv2} demonstrate the 
existence of an extended emission component in addition to the population of bright 
point sources. On the basis of the approach described earlier, we extracted the 
integrated spectrum of both the bright source region (bounded by the spatial mask) 
and the residual emission region (corresponding to the full X-ray extraction region 
less the masked area). 

The relatively large extent and complicated shapes of the bright source and residual 
emission regions makes the evaluation of the background subtraction somewhat more 
complicated than with most \xmmn data. 
In this context, the use of ``blank-sky'' \xmm fields to produce a background spectrum 
would be far from ideal, because the sky X-ray background in the soft band varies from 
field to field, and this is a vital component of the total background signal we wish 
to subtract. Instead, we extracted spectra from both an annulus surrounding the 
defined galaxy
region and also from the corner regions of each detector not exposed to the sky. 
By using an appropriate scaling of these spectra, using the method detailed 
in W07, an appropriate model background spectrum can then be produced.   
The SAS tools {\it arfgen} and {\it rmfgen} were used to produce 
appropriate Auxiliary Response File (ARF) 
and Response Matrix File (RMF) files for the source and residual galaxy regions.
Finally, the counts recorded in adjacent (raw) spectral channels were summed to give 
a minimum of 20 counts per spectral bin in the final set of spectra.
  

\section{Overview of the galactic X-ray properties}
\label{sec:lx}

\subsection{The contribution of luminous point sources}
\label{sec:galaxies}

By application of the spatial mask described earlier, we are able to
measure the count rates (net of background) in the soft, medium and hard 
band images associated with both the ``bright source regions'' and the 
``residual emission regions'' (hereafter we refer to the latter as the 
``residual galaxy'').  Table \ref{table:gal_xray} summarises the results
for each of the five new galaxies in our sample. Here we have converted the 
measured count rates to equivalent fluxes in the band on the basis of a specific 
spectral model and then transposed from flux to luminosity using the distances 
quoted in Table \ref{table:gal:details}. For the bright source regions we assume a 
spectral model consisting of a power-law continuum with photon index $\Gamma=1.7$ 
modified by the foreground absorption in our Galaxy. For the residual
galaxy we use the model which best fits the actual spectrum of
this component, as derived in \S\ref{sec:spec:res}. The quoted X-ray
luminosities (\lx) have been corrected for foreground absorption, that is they 
are unabsorbed values.

The \lx~values reported in Table \ref{table:gal_xray} have been corrected
for two further effects. The first is the spillover of bright source flux into the 
residual galaxy region (discussed earlier).  The second correction relates to the 
fact that the bright source mask obscures a sizeable fraction of the 
galaxy disk. To correct for this, we have scaled up the residual galaxy component 
by a factor derived from interpolating the azimuthally-averaged brightness distribution
(as represented by the radial distributions of the X-ray 
surface brightness discussed in \S \ref{sec:star}) into the masked regions.
The spillover fraction and the area correction factor applicable to 
each galaxy are listed in Table \ref{table:gal_xray}.

The results in Table \ref{table:gal_xray} demonstrate that whereas the set of very 
luminous sources represented by our bright source sample provide the dominant 
contribution to the total galactic X-ray luminosity  above 2 keV, at lower energies, 
and particularly below 1 keV, the split between the bright source and residual 
galaxy contributions is more balanced, albeit with the exception of NGC 300.
For the latter galaxy, we were able to set a much lower threshold for bright 
source removal than was possible for the other galaxies in the sample 
(on account of the relatively proximity of NGC 300 - see Table 
\ref{table:gal:details}). In other words a higher fraction of 
the integrated X-ray flux from point sources was resolved in this 
galaxy.  However, we show later (see \S\ref{sec:star}) that the low surface 
brightness inferred for the underlying diffuse X-ray emission in NGC 300
is mostly likely due to the low SFR density in this Galaxy rather than being
simply a consequence of a more stringent source rejection threshold. 
For M74, NGC 3184, M51 and M83 a broadly similar source detection
threshold transposes to luminosity thresholds for source exclusion
ranging from $2.0 \times10^{37}\ergsec$ (for M83) up to $6.0 \times10^{37}\ergsec$ 
(for NGC 3184). The likely contribution of unresolved sources to
the residual galaxy emission in each case is investigated below and
a correction to a fixed luminosity threshold is applied later in the
analysis (see \S\ref{sec:disc}).

As a further exercise it is possible to estimate the likely contribution 
to the residual galaxy emission of an unresolved population of somewhat less 
luminous sources with spectral characteristics and distribution 
similar to the sources represented in the bright source list.   
Here we make use of recent results from {\it Chandra}, which show that the 
discrete source luminosity function appropriate to the disk regions of spiral 
galaxies typical takes the form of a power-law with a slope in the range 
$-0.5$ to $-0.8$  (in the cumulative form).  For example, \citet{tennant01} 
quote a slope of $-0.5$ for the disk sources in M81, whereas \citet{doane04} 
measure $-0.6$ for NGC 3184 and \citet{pence01} report $-0.8$ for M101. 
\citet{soria03} examine the log 
{\it N} -- log {\it S} relation for relatively faint HMXBs in the disk of M83 
and find a broken power-law provides a reasonable fit to the data, with a low-luminosity 
index of $-0.6$.  Also \citet{colbert04} measure a similar range of power-law 
slopes for the source luminosity function within a relatively large sample of spiral 
galaxies.  The flatness of this form means that relatively small numbers of
very luminous sources provide the bulk of the integrated luminosity residing 
in discrete sources. Furthermore, in the context of providing a rough estimate 
of the contribution of unresolved sources to the residual galaxy signal, it implies 
that we can, in principle, extrapolate below our source detection threshold 
to arbitrarily faint levels (assuming of course that the slope of the luminosity 
function remains constant over an appropriately wide range of source luminosity). 

Here we make the assumption that the slope of the source luminosity function is
$-0.6$ across the set of galaxies and over a wide luminosity range.  We then use 
the number of bright sources actually observed in the galaxy from the threshold 
luminosity up to a fixed upper cut-off (set here as 
\lx $= 5 \times 10^{38}\ergsec$ - see Table \ref{table:gal_sources}), 
to define the normalization of the source luminosity function 
for that particular galaxy.  We first calculate the 
integrated emission of unresolved sources in the broad 0.3-6 keV band and then,
on the basis of the bright source spectral model noted earlier, split this into 
soft-, medium- and hard-band contributions.
The entries in Table \ref{table:gal_xray} under the heading ``unresolved sources'' 
summarises the results on a galaxy by galaxy basis.  In the
soft band, which is the focus of our attention in describing the X-ray morphology
and the linkage of X-ray emission to star formation, the potential contribution of 
unresolved, but relatively luminous, source populations to the measured residual 
galaxy emission ranges from as low as $\sim 10$\% in M51 up to $\sim 40$\% in 
NGC300. By implication the bulk of the residual galaxy luminosity in this band
must reside in some combination of: (i) truly diffuse X-ray emission associated with 
the galaxy disk presumably resulting from the energy input of supernova
and stellar winds and (ii) the integrated emission of populations of relatively 
soft spectrum, lower-luminosity sources such as supernova remnants,  cataclysmic 
variables, active binaries and stellar coronal sources, populations which are 
not well represented in the bright-source ({\it i.e.,} high \lx) samples. 
There is also a potential contribution
from a putative extended galactic halo, which is often seen in edge-on galaxies; 
however the face-on aspect of the present galaxies suggests that a very extended halo 
component will be difficult to distinguish from the components confined to the disk.


\subsection{X-ray characteristics of the individual galaxies}
\label{sec:morphology}

Figures \ref{fig:xuv1} \& \ref{fig:xuv2} compare the galaxy morphology observed 
in soft X-rays with that measured in the far ultraviolet FUV  
($\lambda_{eff} \approx 2267$ \AA) band by {\sl GALEX}\footnote
{The {\sl GALEX} images were obtained from the public archive at http://galex.stsci.edu/GR2/}, 
although in the case of NGC 3184 we use the \xmmn Optical Monitor UVW1 ($\approx 2680$ \AA) 
image recorded in the same observation as the X-ray data (see Table \ref{table:obs})
to make this comparison. In all cases the the UV images were lightly smoothed 
with a circular gaussian mask with $\sigma =$ 4\arcsec.
In a number of the galaxies, particularly M74, M83 and M101, 
the X-ray emission traces segments of the inner spiral arms which are delineated as
very pronounced features in the FUV images.  In all cases the central concentration 
of the soft X-ray emission and its fall-off with increasing galactocentric radius is 
quite well matched to the surface brightness distribution measured in the 
FUV channel. This general correlation of soft X-ray and FUV light
immediately establishes the close connection of a significant component
of the X-ray emission with recent star-formation in these late-type systems. 
We explore the linkage between the X-ray emission and star-formation 
in more detail in \S\ref{sec:star}. Here we briefly comment on each of these 
galaxies, in the context of their extended X-ray emitting components. 

\subsubsection{NGC 300}

NGC300 is an SA(s)d spiral seen at low inclination ($i=30\deg;$ 
\citealt{devaucouleurs91}) at a distance of 2.0 Mpc (\citealt{freedman01}). 
Many large HII regions are visible throughout the galaxy, and there is evidence 
that the galaxy has experienced many episodes of star formation. However, star 
formation in the central regions appears to be suppressed in recent times, 
with a small population of stars present in the central arcminute of the galaxy 
with ages below 1 Gyr (\citealt{davidge98}). There is clear evidence of spiral structure, 
although the contrast between ``arm'' and ``off-arm'' regions in the galaxy is low. 
In X-rays NGC300 has been studied by \rosat (\citealt{read01}) and more recently by 
\xmmn (\citealt{carpano05}). \citet{carpano05} report
the presence of diffuse emission in the galactic disk of NGC 300 with 
\lx $<10^{38}\ergsec$  and temperatures of $\approx 0.2$ 
and $\approx 0.8$ keV.
In the current paper we determine the aggregate luminosity of the 20 brightest 
point sources within the central 5$\arcm$ of NGC300 to be \lx $\approx 5 \times 10^{38} 
\ergsec$, in agreement with earlier studies.  We estimate the 
X-ray luminosity of the residual emission to be $\approx 5 \times 10^{37} 
\ergsec$ (0.3--1 keV). The surface brightness of this component is extremely low 
but appears to trace a central bar-like structure and to
extend out to at least a distance of 2.5 kpc ($4\arcm$).

\subsubsection{M74}

M74 (NGC 628) is an SA(s)c spiral seen almost face-on ($i=7\deg;$ \citealt{shostak84}) 
at a distance of 11 Mpc (\citealt{gildepaz07}). The nuclear region of the galaxy 
is characterised by strong HII emission with no evidence for the presence of an AGN. 
A grand-design twin-armed spiral structure is observed. In X-rays M74 has been 
studied by both \chandra (\citealt{krauss05}) and \xmmn (\citealt{soria02a}; 
\citealt{soria04}), albeit with little attention, to date, on the 
diffuse X-ray component.
In the present analysis we confirm the presence of an extensive discrete source 
population in M74 with an aggregate luminosity in the 20 brightest point 
sources amounting to $5.7 \times 10^{39} \ergsec$.  The underlying residual
X-ray emission has a luminosity of $2.1 \times 10^{39} \ergsec$. This component
traces the inner spiral arm structure and is evident out to a 
distance of 9 kpc (3$\arcm$), at which point it falls below the background level.

\subsubsection{NGC 3184}

NGC3184 is an SAB(rs)cd barred spiral seen at low inclination ($i<24\deg$; Lyon-Meudon 
Extragalactic Database [LEDA]\footnote{at http://leda.univ-lyon1.fr}) at 
a distance of 11.6 Mpc 
(\citealt{leonard02}). The nuclear region of the galaxy is dominated by HII 
emission with no compelling evidence for an AGN (\citealt{doane04}). Grand-design 
twin spiral arms are observed, although the contrast in brightness between ``arm'' 
and ``off-arm'' regions in the galaxy is low. 
In X-rays NGC3184 has been studied by \chandra with an emphasis on both the point 
source population (\citealt{colbert04}) and the diffuse X-ray emitting gas 
(\citealt{doane04}; \citealt{tyler04}). 
The diffuse X-ray emission shows a strong correlation with regions which are bright
in \Ha~(\citealt{tyler04}). \citet{doane04} report the presence of 
diffuse X-ray emission concentrated in areas of younger stellar populations and 
star-forming regions, with a surface brightness five times greater in spiral arm regions 
than in off-arm regions. The spectrum of the diffuse emission in the 
galactic disk can be well fitted with a two-temperature thermal model with kT of 
$0.13$ keV and $0.43$ keV (\citealt{doane04}).
In the present study we measure the aggregate luminosity of 
the 18 brightest point sources to be $\approx 6 \times 10^{39} \ergsec$, 
consistent with earlier \chandra estimates (\citealt{colbert04}).
The residual X-ray emission has an X-ray luminosity of $2.5 \times 10^{39} \ergsec$
and extends from the nucleus out to a distance of 8 kpc ($2.5\arcm$). 

\subsubsection{M51}

M51 (NGC5194) is an SA(s)bc spiral seen almost face-on ($i=20\deg$;
\citealt{tully74}) at a distance of 8.4 Mpc (\citealt{feldmeier97}). 
It hosts a low-luminosity active galactic nucleus, which shows optical emission 
lines and is classified as a Seyfert 2 (\citealt{stauffer82}; \citealt{ho97}). 
M51 is interacting tidally with a companion SB0/a galaxy (NGC5195), and both systems 
are seen to contain starburst activity. There is grand-design twin-arm spiral structure 
present. In X-rays M51 has been studied by \einstein (\citealt{palumbo85}), 
\rosat (\citealt{marston95}; \citealt{ehle95}), \asca (\citealt{terashima98}) and 
more recently by \chandra (\citealt{terashima04}; \citealt{colbert04}; \citealt{tyler04}) 
and \xmmn (\citealt{dewangan05}). Studies have also been carried out at higher 
X-ray energies by \bepposax (\citealt{fukazawa01}). 
The diffuse X-ray emission is well correlated with the radio continuum
and mid-infrared emission distributed in the spiral arms establishing a
clear link with star formation (\citealt{ehle95}; \citealt{tyler04}). 
\citet{terashima04} find that the diffuse X-ray emission in the southern extranuclear 
cloud is well fitted with a single-temperature thermal model with kT of $0.58$ keV.  
The current analysis confirms 
the general properties of the galaxy established by the \rosat and \chandra observations. 
The aggregate luminosity of the brightest 25 point sources in M51 
is $1.7 \times 10^{40} \ergsec$ (0.3--6 keV), 28\% of which is attributable
to the nuclear source.
Our analysis reveals residual soft
X-ray emission extending from the bright nucleus along the spiral arms out to a 
distance of 5 kpc ($2.5\arcm$). Lower surface brightness emission from the extended 
disk is observed out to a radial distance of 9 kpc in all directions, with further 
X-ray emission joining M51 to its companion galaxy.  The X-ray luminosity
of the residual component is  $7.2 \times 10^{39} \ergsec$.

\subsubsection{M83}

M83 (NGC 5236) is an SAB(s)c barred spiral seen at low inclination ($i=24\deg$; 
\citealt{talbot79}) 
at a distance of 4.5 Mpc (\citealt{thim03}). It harbours an active
nuclear region undergoing a violent starburst, a circumnuclear starburst 
and grand-design twin-armed spiral structure.  There are numerous sites of 
active star-formation coincident with the spiral arms (\citealt{vogler05}). In X-rays 
M83 has been studied by \einstein (\citealt{trinchieri84}), 
\rosat (\citealt{ehle98}; \citealt{immler99}), \asca(\citealt{okada97} and 
more recently by \chandra (\citealt{soria02}; \citealt{soria03}; \citealt{colbert04}; 
\citealt{tyler04}). 
On an arcminute scale
both the \rosat and \chandra X-ray observations trace apparently diffuse
emission associated with the inner spiral-arms
(\citealt{immler99}; \citealt{soria03}; \citealt{tyler04}).  \citet {immler99} 
report the diffuse emission associated with a cluster of bright HII regions 
embedded in the south-western spiral arm as having a two-temperature thermal 
characteristic with kT of $0.26$ keV and $0.95$ keV. 
In the present analysis we measure the aggregate luminosity of the bright point 
sources in M83 to be $L_X = 7.6 \times 10^{39} \ergsec$, which includes 
a contribution of $3.1 \times 10^{39} \ergsec$ from the central 
region.  
The \xmmn soft band image traces emission from the bright nuclear region
out along the twin spiral arms to a galactocentric distance
of about 4 kpc ($3\arcm$). Extended X-ray emission of relatively low
surface brightness can then be further traced out to $\sim 8$ kpc. 
The X-ray luminosity of the residual component is  $4 \times 10^{39} 
\ergsec$.


\section{Spectral Analysis}
\label{sec:spectrum}

Using the methodology outlined in \S\ref{sec:diff_extr}, spectral datasets 
were obtained for both the bright-source and the residual-galaxy regions
for four galaxies, namely M74, NGC3184, M51 and M83. 
For simplicity, we based our analysis on a single observation for each galaxy, 
namely the observation with the deepest pn exposure (see Table \ref{table:obs}). 
Due to limited signal-to-noise in the MOS channels, we focus on the spectra 
derived from the EPIC pn camera, except in the case of M83, where both pn and 
MOS spectra were fully utilised and in M74 where both datasets were employed 
in the spectral investigation of its bright-source region.
Unfortunately the residual-galaxy emission in NGC 300 was too faint for 
detailed spectral analysis to be merited.  The spectral fitting was carried 
out using the software package XSPEC Version 12.4.

\subsection{Spectra of the bright-source regions}
\label{sec:spec:sources}

\begin{figure*}
\centering
\rotatebox{270}{\scalebox{0.4}{\includegraphics{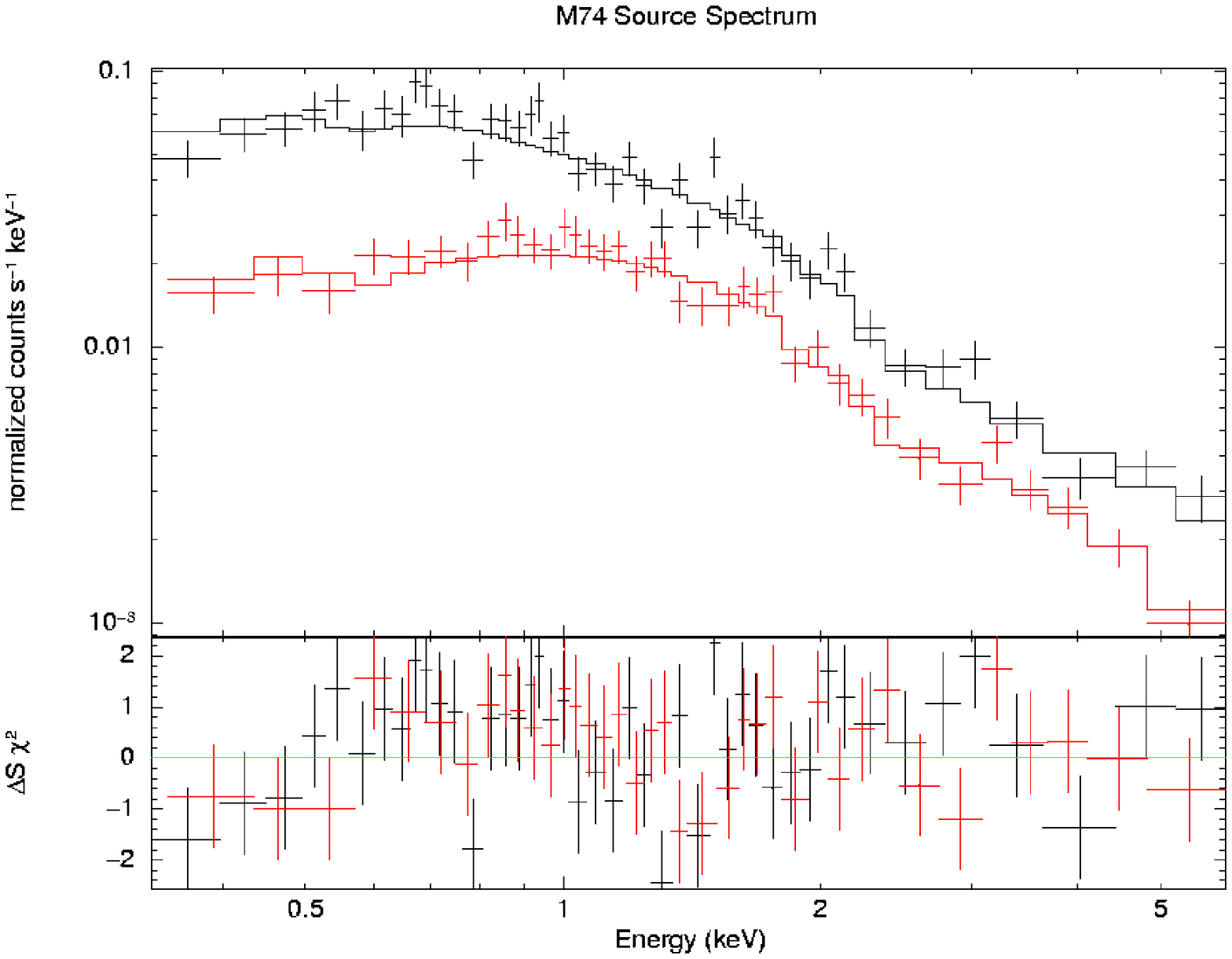}}}
\rotatebox{270}{\scalebox{0.4}{\includegraphics{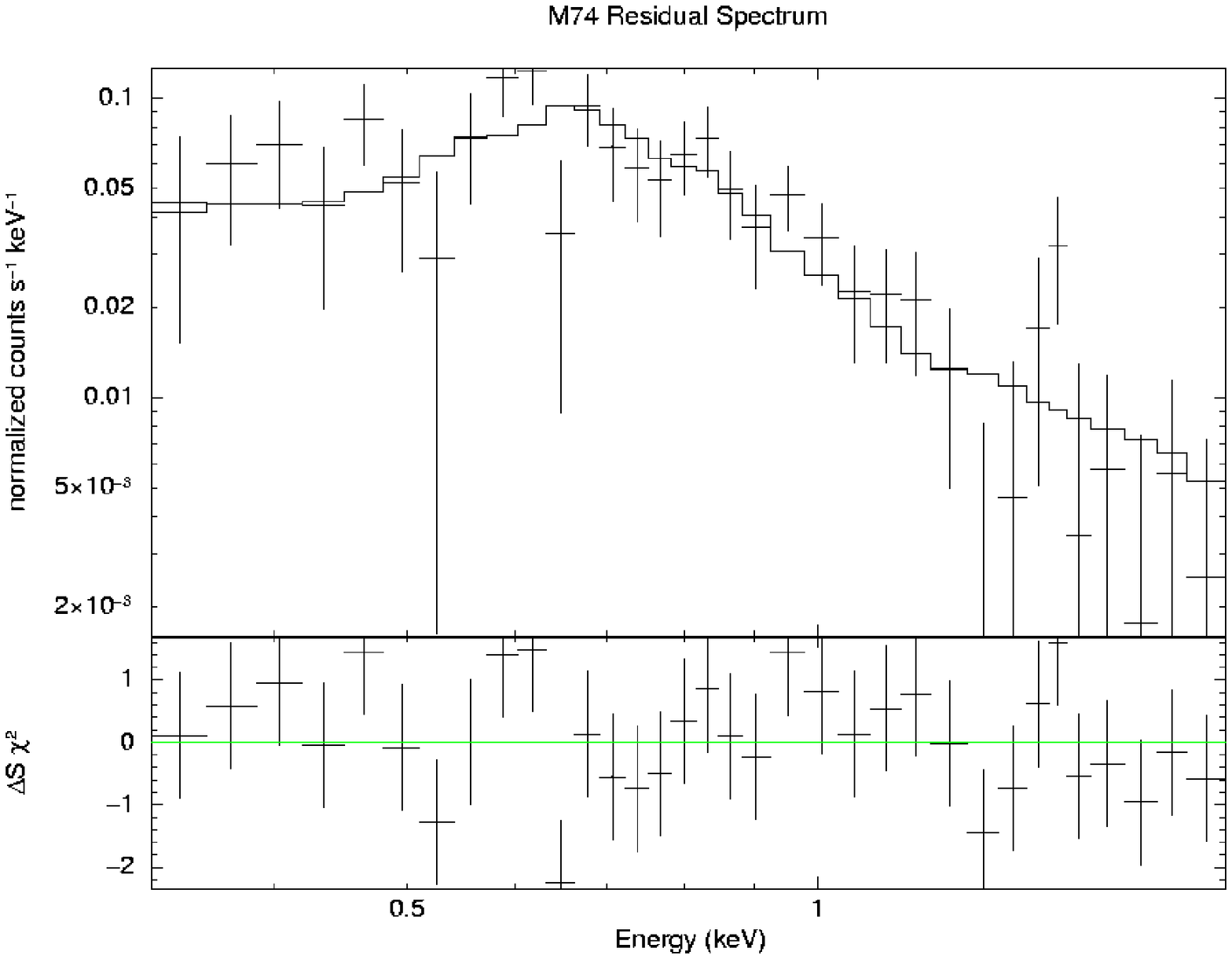}}}
\rotatebox{270}{\scalebox{0.4}{\includegraphics{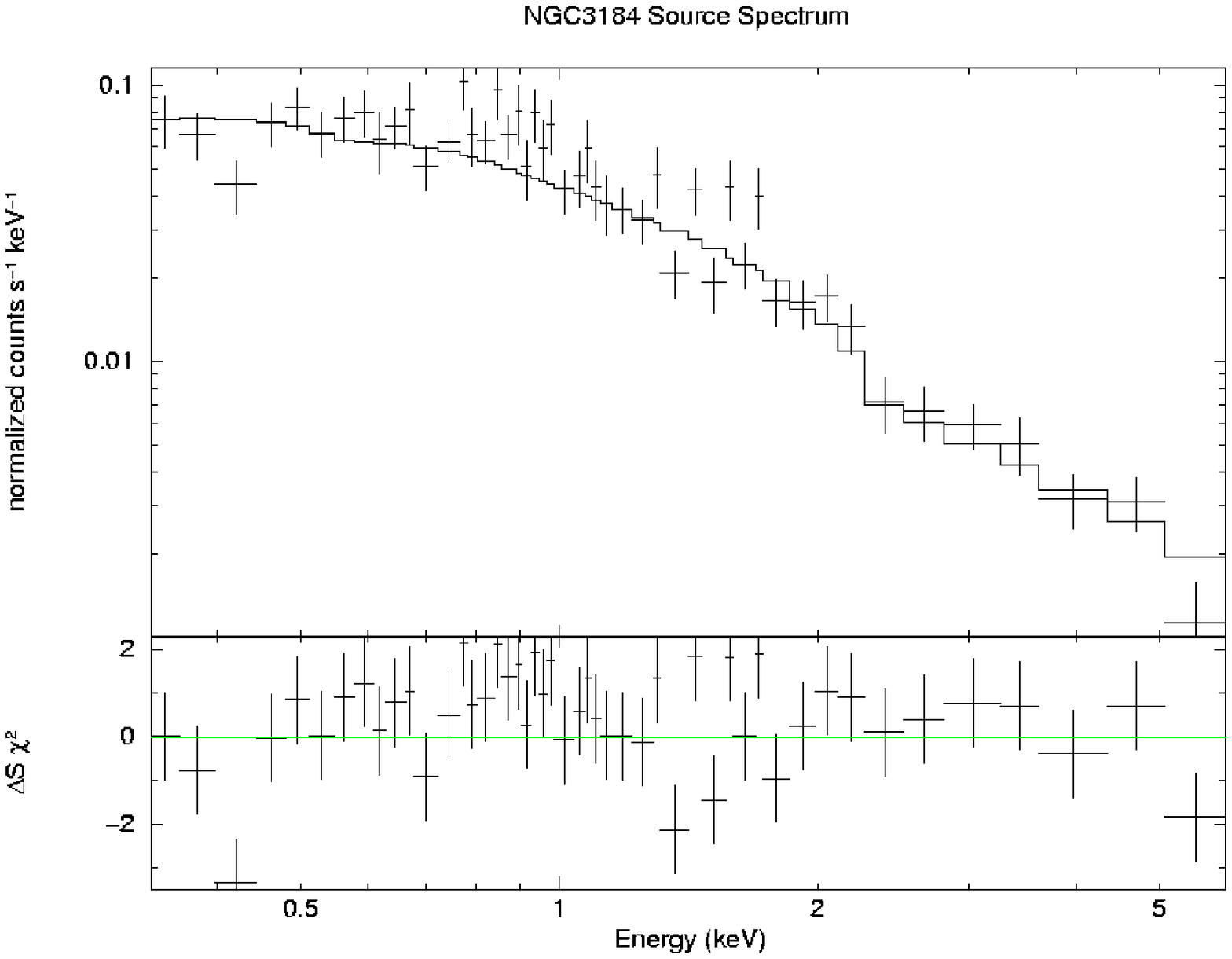}}}
\rotatebox{270}{\scalebox{0.4}{\includegraphics{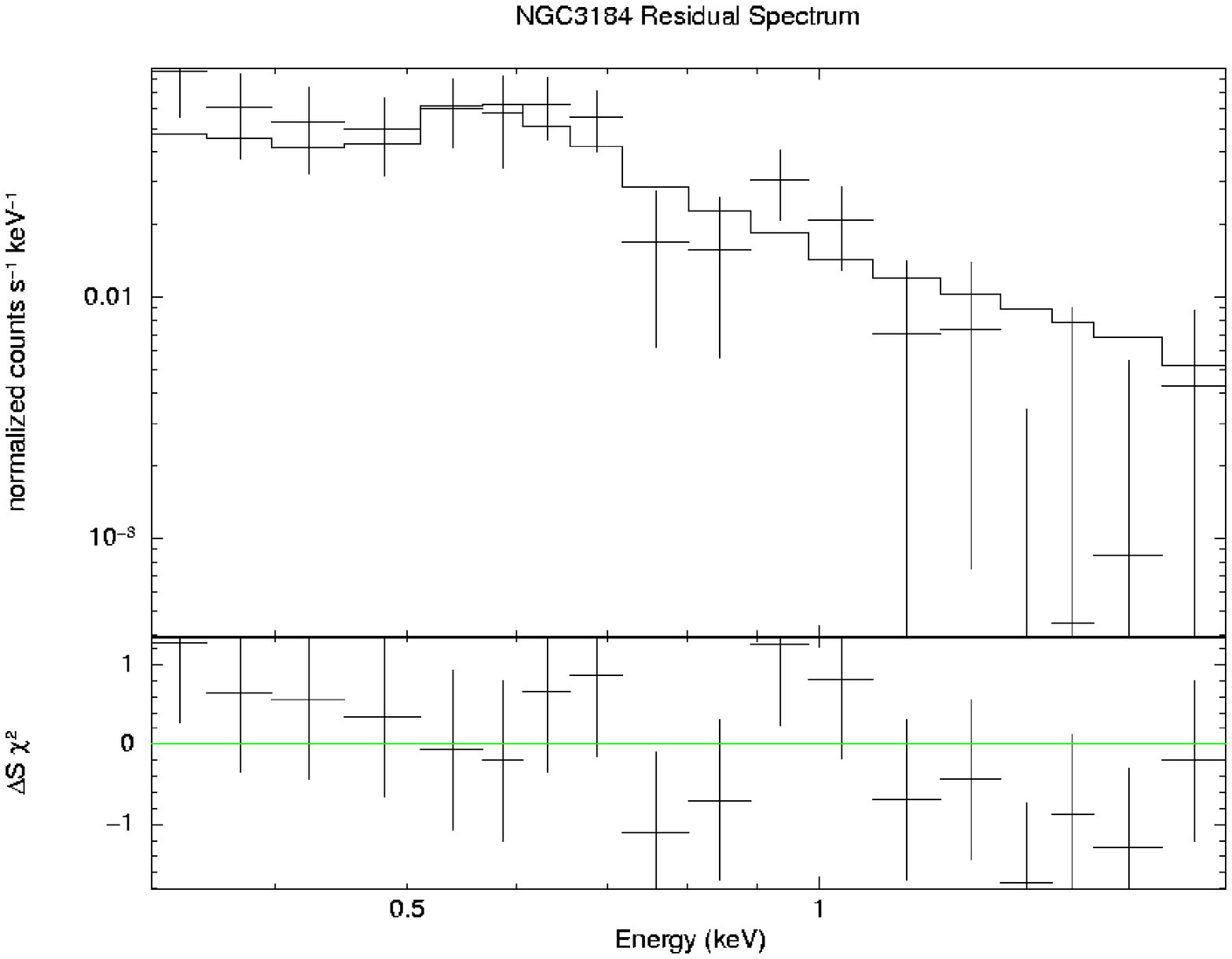}}}
\rotatebox{270}{\scalebox{0.4}{\includegraphics{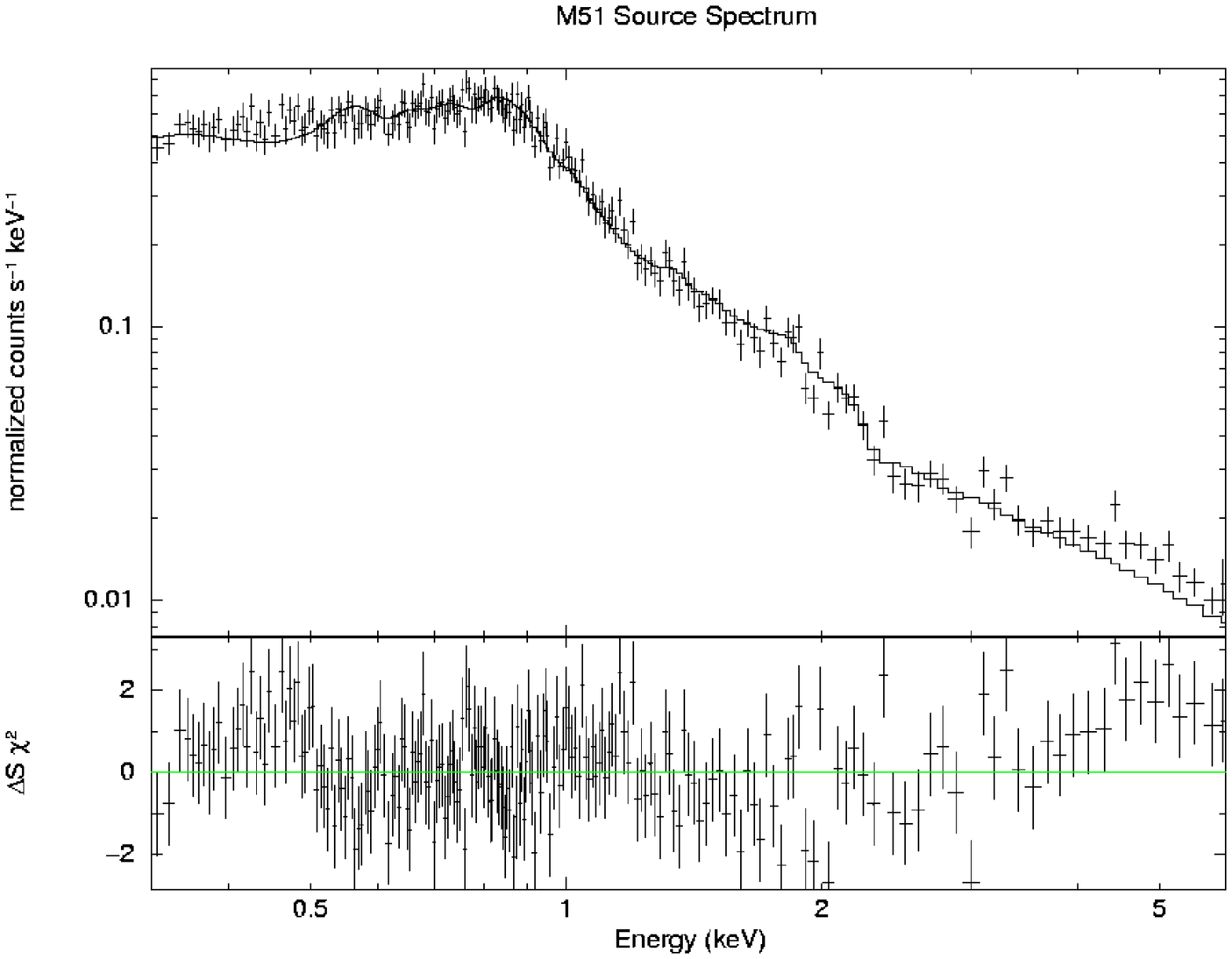}}}
\rotatebox{270}{\scalebox{0.4}{\includegraphics{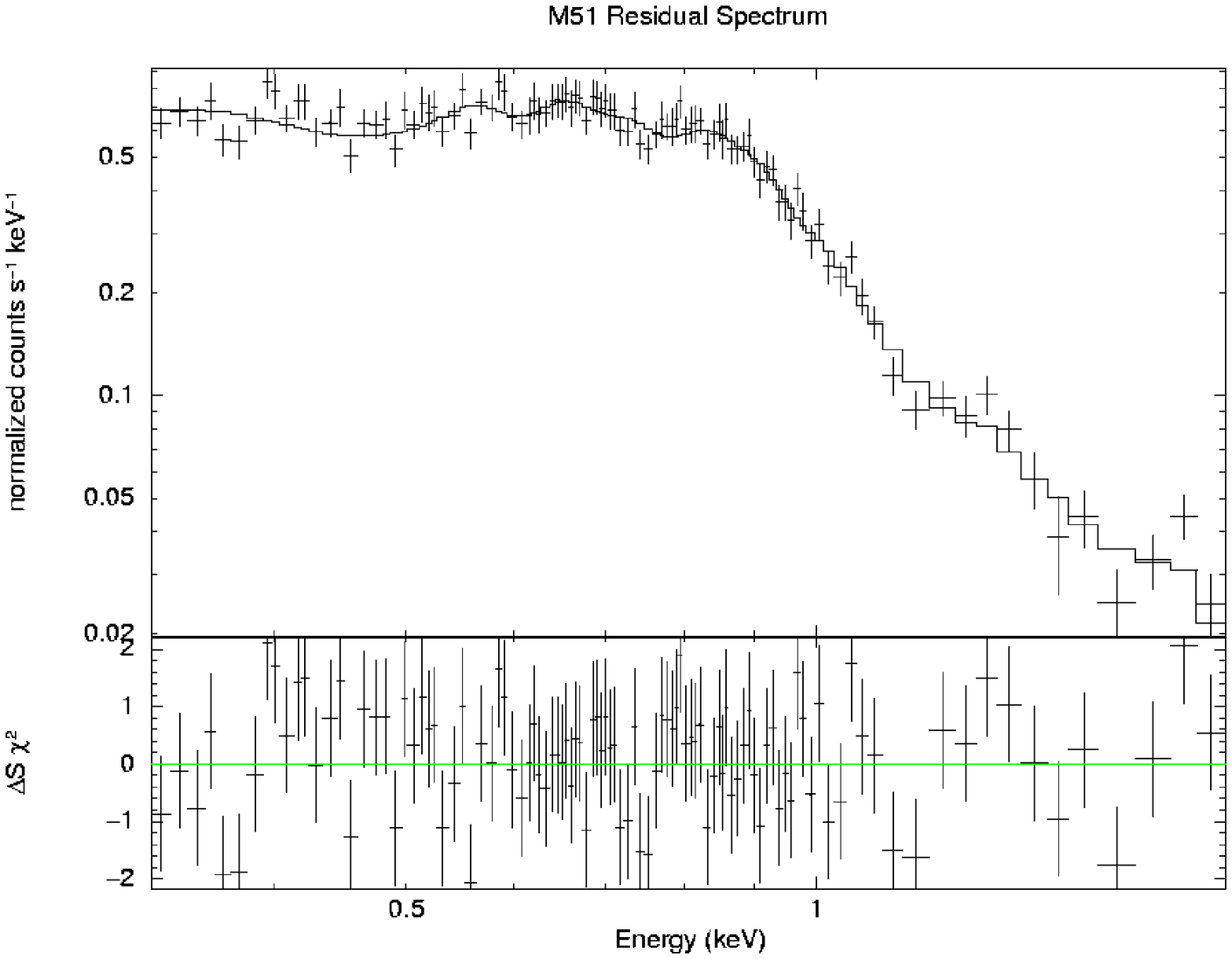}}}
\rotatebox{270}{\scalebox{0.4}{\includegraphics{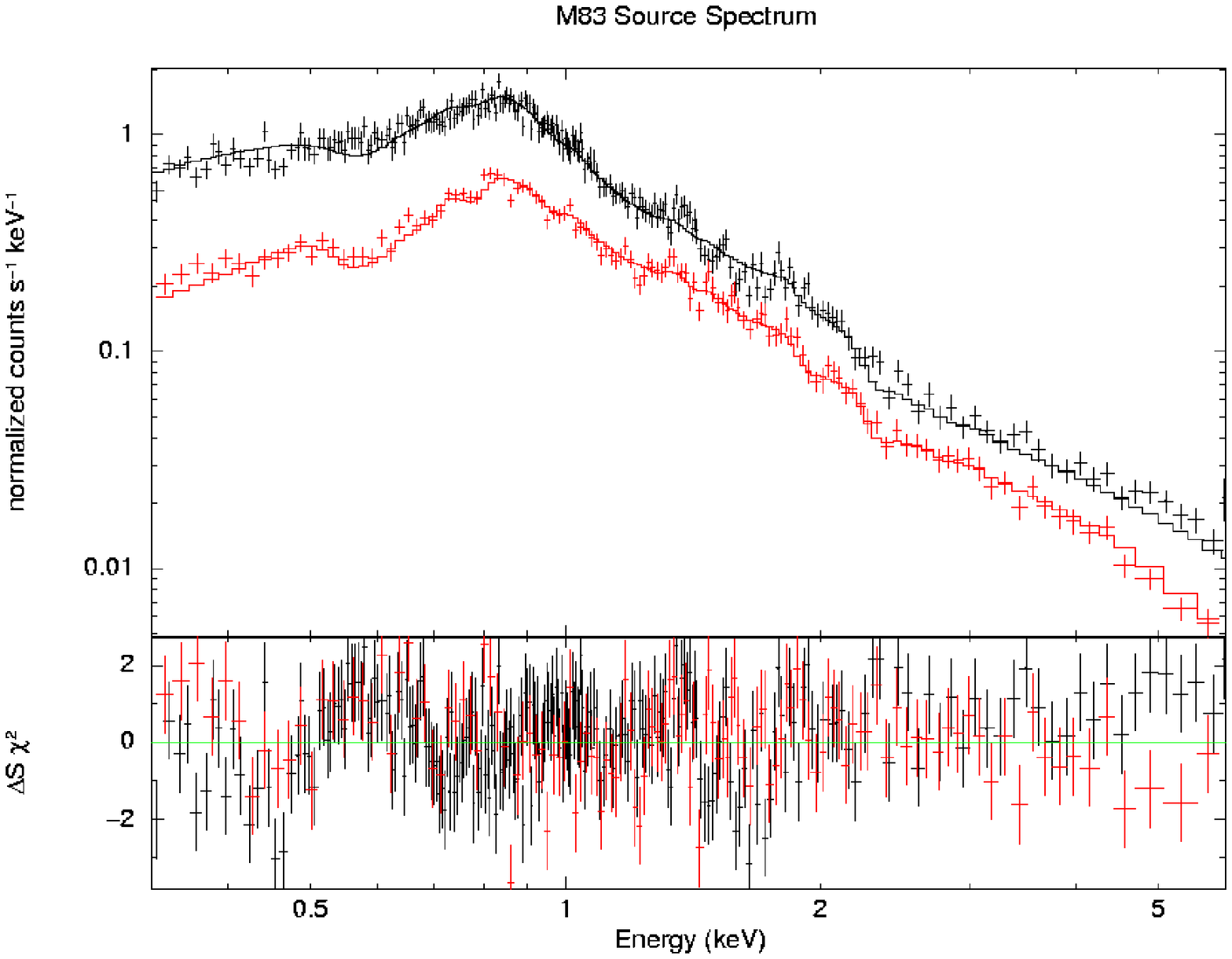}}}
\rotatebox{270}{\scalebox{0.4}{\includegraphics{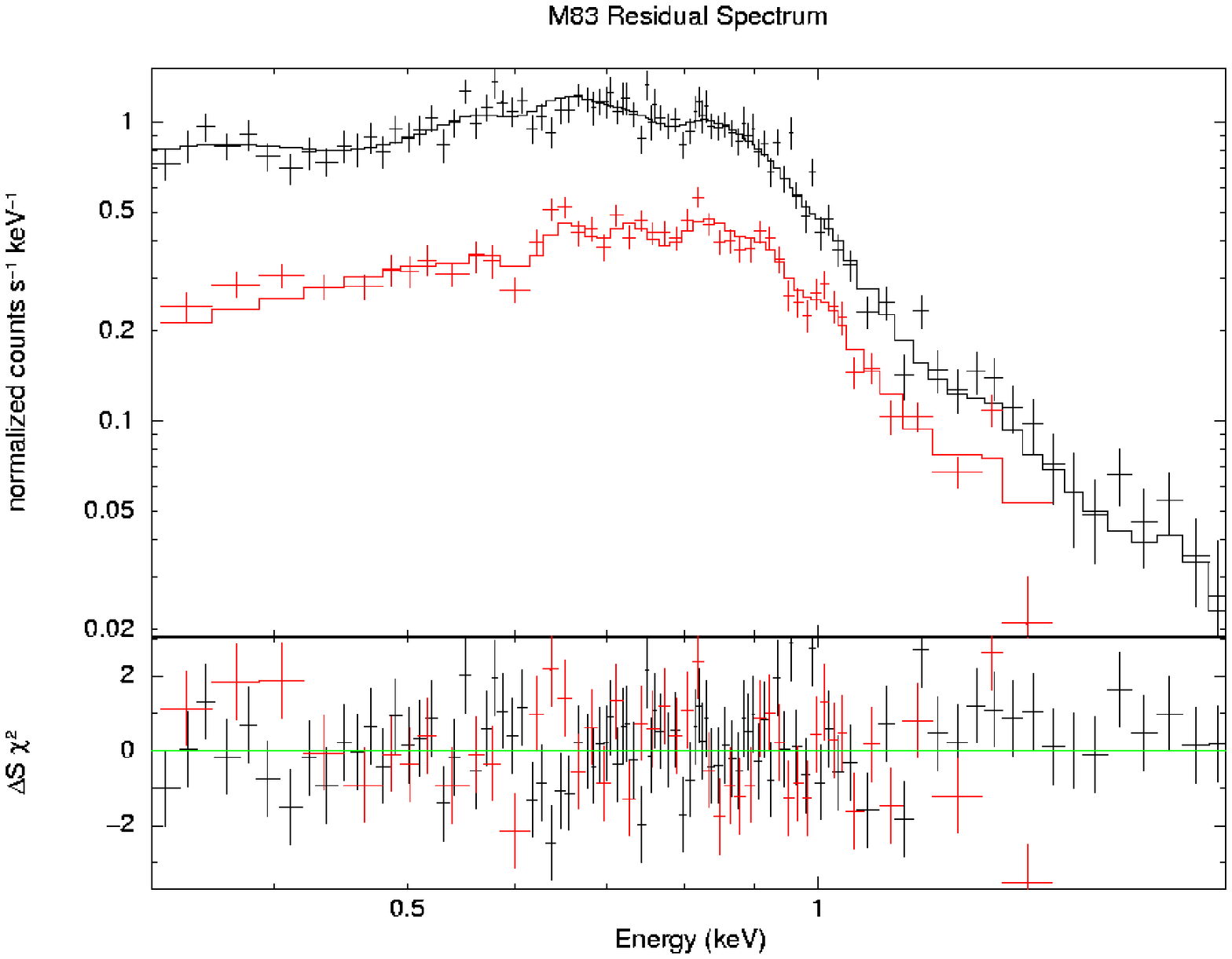}}}

\caption{The EPIC spectra for four galaxies. {\it Left-hand panels:} 
The spectra of the bright-source regions. {\it Right-hand panels:} The spectra of
the residual-galaxy regions.  In all cases the solid line corresponds 
to the best-fit spectral model (see text). The $\chi^{2}$ residuals with respect 
to the best-fitting model are also 
shown. }

\label{fig:spec}
\end{figure*}

The spectra representative of the bright-source regions in M74, NGC3184, M51 and M83
were well fitted with either a simple power-law continuum or a power-law continuum 
plus an additional solar-abundance thermal plasma component (the MEKAL
model in XSPEC). The best-fit photon indices were all in the range 1.55--1.9 with 
the temperature of the thermal emission typically $\approx 0.65$ keV. 
Fig. \ref{fig:spec} shows the measured spectra, the best-fitting 
model and residual $\chi^{2}$ to these best fits. Similarly, 
Table \ref{table:spec_fits} 
summarises the details of the best-fitting models. After correcting for the flux 
spillover beyond the spatial mask, the inferred luminosities in the integrated
bright-source spectra were found to be in general agreement with the estimates 
quoted previously based on the image analysis.
The photon index observed is typical of the spectra of HMXB and LMXB found in the 
Local Group (\citealt{grimm07}). The thermal emission 
associated with the bright-source regions is of similar temperature to the ``hot''
component observed in the residual-galaxy emission (see below). However, where this
component is detected in the bright-source spectra, namely in NGC3184, M51 and M83, 
it appears at a proportionately higher level than might be inferred based on a simple 
area scaling with respect to the residual-galaxy emission.


\begin{table*}
\caption{Parameters of the best-fitting models to the spectra of the bright-source and 
residual-galaxy regions.}
\centering
\begin{tabular}{lccccccc}
\hline

Galaxy  & Region  & Power-law Cont. & Cool MEKAL & Hot MEKAL  &  Goodness  
&  Cool:Hot \\
& &  Photon Index  &    keV  & keV  & of Fit & Flux Ratio \\
& &  Normalization  &  Normalization  &  Normalization  &  $\chi^{2}$/dof  & (0.3--2 keV) \\
\hline
M74 & Bright Sources & 1.64$\pm$0.06  &  -  &  -  &  146/150  &  -  \\
&  &  $5.62 \times 10^{-5}$  &    &    &    &    &    \\
\\
& Residual Galaxy & -  &  0.26$\pm$0.04  &  [$0.65$]$^{a}$  &  197/202  &  $> 5.4$  \\
&  &    &  $3.74 \times 10^{-5}$  &  [$< 5.5 \times 10^{-6}$]$^{a}$ &   &    &    \\
\\
NGC3184 & Bright Sources  &  1.57$\pm$0.10  &  -  &  0.71$\pm$0.13  &  86/78  &  -   \\
&  &  $4.41 \times 10^{-5}$ &    &  $6.63 \times 10^{-6}$  &   &   &   \\
\\
&  Residual Galaxy  &  -  &  0.20$\pm$0.06  &  [$0.65$]$^{a}$  &  122/105  &  $> 4.1$ \\
&  &    &  $1.85 \times 10^{-5}$  &  [$< 3.5 \times 10^{-6}$]$^{a}$  &    &    &    \\  
\\
M51 & Bright  Sources  &  1.57$\pm$0.05  &  0.19$\pm$0.02  &  0.61$\pm$0.05  &  389/378 &  -  \\
&  &  $2.65 \times 10^{-4}$  &  $1.20 \times 10^{-4}$  &  $1.48 \times 10^{-4}$  &    &    &    \\
\\
&  Residual Galaxy &  2.90$\pm$0.15  &  0.24$\pm0.02$  &  0.64$\pm0.04$  &  210/260  &  1.3 \\
&  &  $1.14 \times 10^{-4}$  &  $1.63 \times 10^{-4}$  &  $9.94 \times 10^{-5}$  &    &    &    \\
\\
M83 & Bright Sources  &  1.86$\pm$0.02  &  -  &  0.61$\pm$0.03  &  796/679  &  -   \\
&  &  $5.42 \times 10^{-4}$  &    &  $2.78 \times 10^{-4}$  &    &  &   \\
\\
&  Residual Galaxy &  2.92$\pm$0.15  &  0.24$\pm$0.02  &  0.64$\pm$0.03  &  404/356  &  1.5  \\
&  &  $1.12 \times 10^{-4}$  &  $2.32 \times 10^{-4}$  &  $1.19 \times 10^{-4}$  &    &    &    \\
\hline    
\end{tabular}
\\
$^{a}$ - Upper bound on the normalization of a 0.65-keV thermal component (95\% confidence interval). \\
\label{table:spec_fits}
\end{table*}


\begin{table*}
\caption{Physical properties of the diffuse gas present in each galaxy.}
\centering
\begin{tabular}{lcccccccc}
\hline

Galaxy  &  Radius$^{a}$  &  Component   & Electron Density  &  Thermal Energy  &  Cooling Timescale   \\
& kpc  &  (keV)  &  $10^{-3} \eta^{-1/2}\rm ~cm^{-3}$  &  $10^{55} \eta^{1/2} erg$  &  $10^{8}\eta^{1/2}$ yr \\
\hline
\\
M74 &  10.0  &  0.2  &  3.1  &  2.5  &  6.6 \\
\\
NGC3184 &  10.0  &  0.2  &  2.1  &  1.9  &  4.3 \\
\\
M51 &  7.5  &  0.2  &  6.7  &  2.4  &  1.9 \\
&  &  0.65  &  5.2  &  5.5  &  8.3 \\
\\
M83 &  6.5  &  0.2  &  4.9  &  2.5  &  4.2 \\
  &  &  0.65  &  3.1  &  3.6  &  9.5 \\
\hline    
\end{tabular}
\\
$^{a}$ - Assumed radius of a putative shallow halo component (see text)\\
\label{table:gas_physics}
\end{table*}



\subsection{Spectra of the residual-galaxy regions}
\label{sec:spec:res}

Previously we estimated that 9-12 per cent of the counts from the bright sources 
overspill 
into the residual-galaxy regions. In carrying out the spectral analysis of the latter
we have correct for this spillover effect, by including an appropriate fraction 
of the best-fit bright-source model as a fixed component of the spectral
model.  Visual inspection of the residual-galaxy spectra revealed almost no emission 
above 1.5 keV in any of the galaxies.  It was further found that thermal models 
comprising either one or two solar-abundance\footnote{In general the quality of the
spectra were insufficient to allow useful contraints to be placed on the metal 
abundances
in the thermal plasmas. As noted in W07, the fitting of low resolution spectra 
pertaining to a complex multi-temperature plasma with simple models
can often result in the artifical requirement for strongly subsolar abundances.}
MEKAL components provided
a reasonable fit to the residual-galaxy spectra.  In all cases a ``cool'' thermal
component was required at kT $\approx0.2$ keV. The spectra with the best signal-to-noise 
ratio, namely those for M51 and M83, also
required a ``hot'' thermal component at kT $\approx 0.65$ keV.
These spectral characteristics appear to be quite typical of the diffuse components seen 
in normal and starburst galaxies ({\it e.g.,}, 
\citealt{fraternali02}) and consistent with previously published measurements for 
several of the galaxies in the sample ({\it e.g.,} \citealt{ehle98}; 
\citealt{soria03}; \citealt{kuntz03}; \citealt{carpano05}; W07).
The spectra of M51 and M83 also showed evidence for a ``soft excess'' which, 
via spectral fitting, was modelled as  a soft power-law component with 
photon index $\approx 2.9$. 
This latter component contributes 26\% of the flux in the 0.3-2 keV band 
for M83 and 35\% for M51. This soft excess may well relate to the
contribution of populations of soft sources most notably recent supernova
remnants and, in fact, there is ample observational evidence that such sources
are very often prominent in the spiral arms of late-type galaxies, for example
IC342 (\citealt{kong03}), M33 (\citealt{pietsch04}; \citealt{grimm05}) and M83 
(\citealt{soria03}).  After applying an appropriate area-correction factor,
the luminosities derived from the spectra data were broadly consistent with those
derived from the imaging analysis (see Table \ref{table:gal_xray}).

There are two points pertinent to the above spectral analysis. 
The first is that the spectral modelling did not require the inclusion of 
any absorption over and above the foreground Galactic $N_{H}$ which was included as
a fixed component in all the models. The implication is that any absorption
intrinsic to the host galaxy must be relatively low, commensurate with the fact 
that we are studying systems which have a near face-on aspect. The second point is that
there was no strong evidence in the residual galaxy spectra
for a ``hard-excess'' component attributable to an unresolved population
of relatively hard discrete sources. This confirms the conclusions
in \S\ref{sec:galaxies}, namely that relatively luminous binaries (albeit below
the applied luminosity threshold) do not provide a substantial contribution to
of the residual-galaxy emission.
In fact the lack of clear evidence for hard emission in the spectra of the
residual galaxy regions (and also in the corresponding hard-band images) suggests
that the estimates in \S\ref{sec:galaxies} of the potential contribution of 
relatively luminous discrete sources might be best considered as upper-limit values. 

For each of the galaxies in the sample, we have measured the relative contribution 
of the 0.2-keV and 0.65-keV MEKAL components to the X-ray flux  measured in the
0.3--2 keV band. The results are summarised in Table \ref{table:spec_fits}, where
we include the upper limits derived for a fixed 0.65-keV component in the
case of M74 and NGC 3184.  For comparison, the corresponding flux ratio
pertaining to the cool and hot MEKAL components in M101 (from the spectral 
fitting results in W07) is $\approx 3.8$. 

If we assume that all of the emission attributed to thermal components
originates in truly diffuse plasma, then it is possible to infer some 
physical properties of the medium on the basis of a particular
geometrical configuration. Here we assume that the thermal plasma 
is contained within a cylindrical region extending from the centre of the galaxy out 
to $75\%$ of the \d25 radius
with a half-width perpendicular to the plane of 
0.5 kpc, representing a shallow halo. Using this volume, we can infer the mean 
electron density $n_{e}$ through the derived emission measure $\eta n^{2}_{e} V$ 
(where $\eta$ is the `filling factor' - the fraction of the total volume $V$ 
which is occupied by the emitting gas). The derived density combined with
the volume and temperature then leads to an estimate of the thermal energy 
contained within the gaseous component. Based on the measurements from the 
spectral fitting, we derive the results summarised in Table \ref{table:gas_physics}.  
In the cases of M51 and M83, where the balance between the cool and hot thermal
components is most strongly weighted towards the hot component, we find that the 
thermal energy contained in the hotter component is significantly greater 
than that in the cooler component and that these two components 
are not in pressure balance.  Our spectral modelling of the residual emission
in  NGC3184 and M74, although not requiring a 0.65-keV spectral component, does 
not exclude the presence of such a component in rough pressure balance with the 
cooler plasma, consistent with the situation pertaining in M101 (W07).  
The cooling of these plasmas is dominated by line emission with radiative
timescales in the range $2-10 \times 10^{8} \eta^{1/2}$\,yr 
(Table \ref{table:gas_physics}).

Of course, as noted previously, a more likely scenario is that the bulk of the
observed soft emission originates in the combination of diffuse emission 
closely associated with star-formation in the galactic disk and the integrated
emission of the populations of soft sources within the disk. 


\section{The connection between soft X-ray emission and star-formation}
\label{sec:star}

As noted earlier, the general similarity of soft X-ray and FUV images establishes 
a close linkage between the X-ray emission and recent star-formation in these 
late-type systems. This connection has previously been demonstrated in many 
individual galaxies and also in samples of objects; for example \citet{tyler04}
find a very strong correlation between X-ray emission and both \Ha~and 
mid-infrared emission in a set of spiral galaxies observed with
{\it Chandra}. In our early study of M101 (W07), we further investigated how 
the overall point-to-point correlation varies when one contrasts the soft X-ray 
surface brightness with that measured in different wavebands extending from the FUV
through to optical V band. The fact that the best match was between the soft 
X-ray and U band images was interpreted in terms of two spatial components, namely 
a clumpy thin-disk component which serves as a tracer of the spiral arms plus 
a more extended component, possibly located in the lower-halo with a larger filling 
factor, which produces the central concentration of the soft X-ray emission (W07).
However, in the present paper we focus on the relationship between the residual 
galaxy X-ray emission and the corresponding FUV emission in terms of their 
radial-averaged spatial distribution.

\begin{figure*}
\centering
\rotatebox{270}{\scalebox{0.4}{\includegraphics{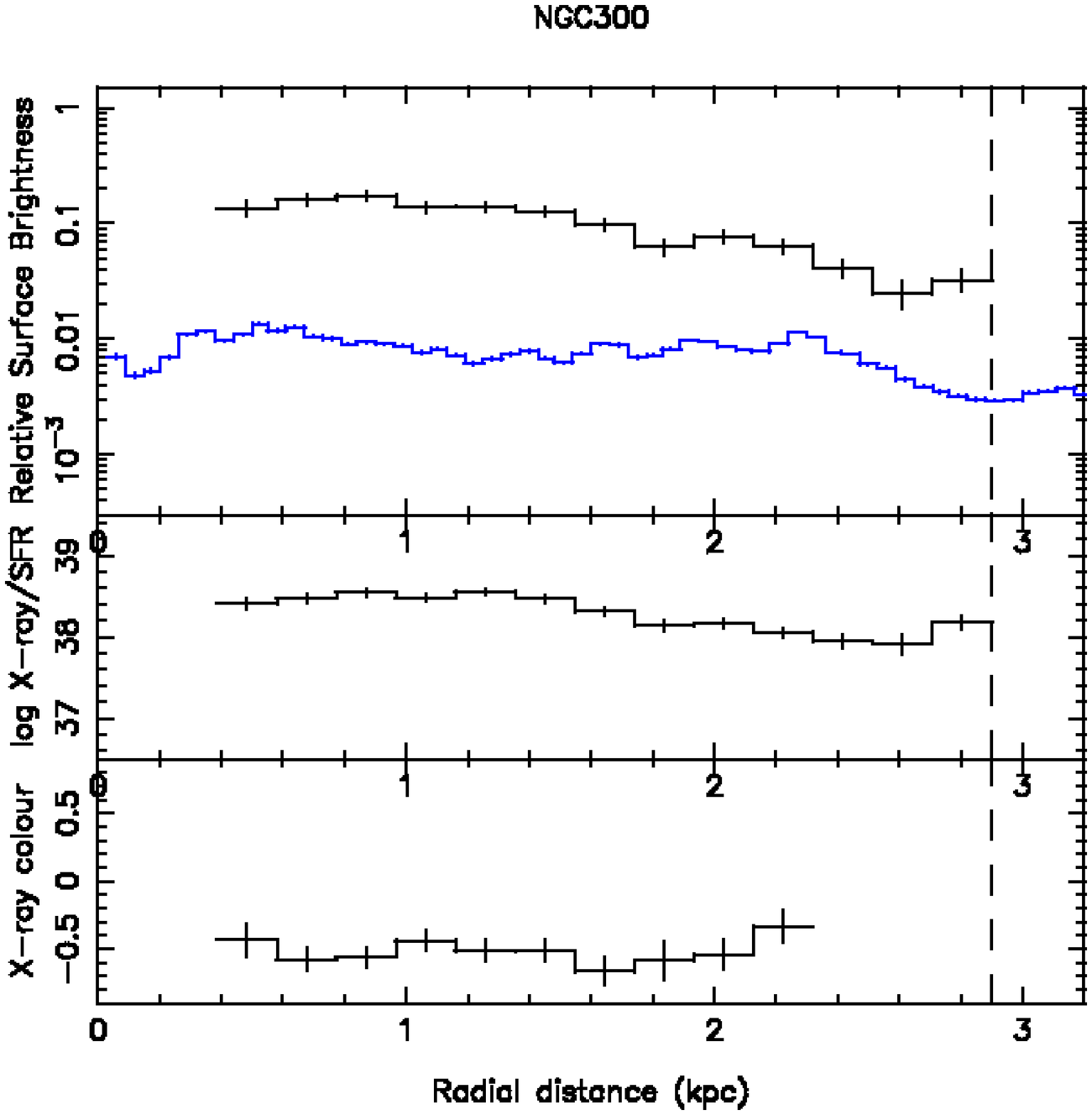}}}
\rotatebox{270}{\scalebox{0.4}{\includegraphics{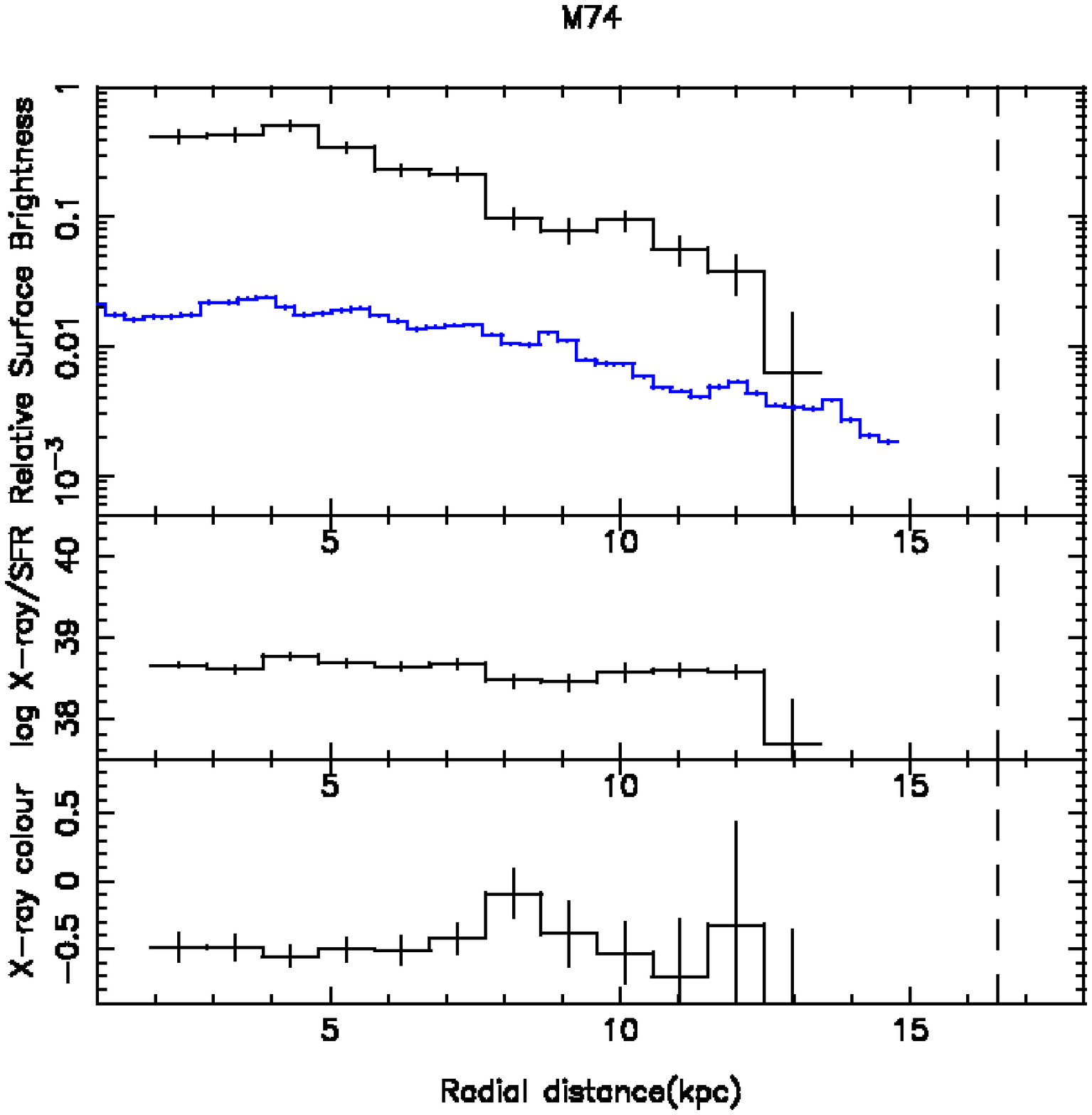}}}
\rotatebox{270}{\scalebox{0.4}{\includegraphics{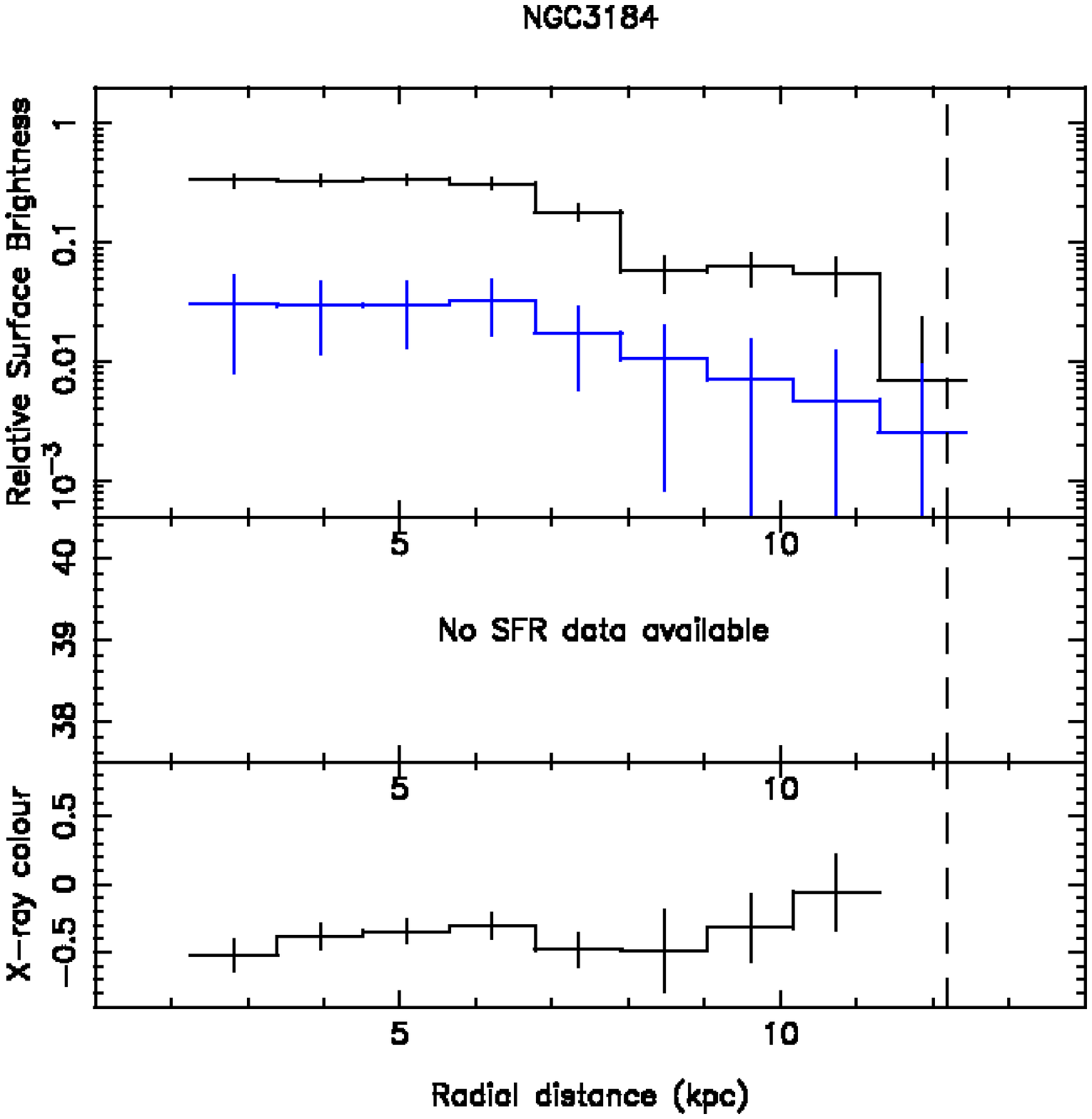}}}
\rotatebox{270}{\scalebox{0.4}{\includegraphics{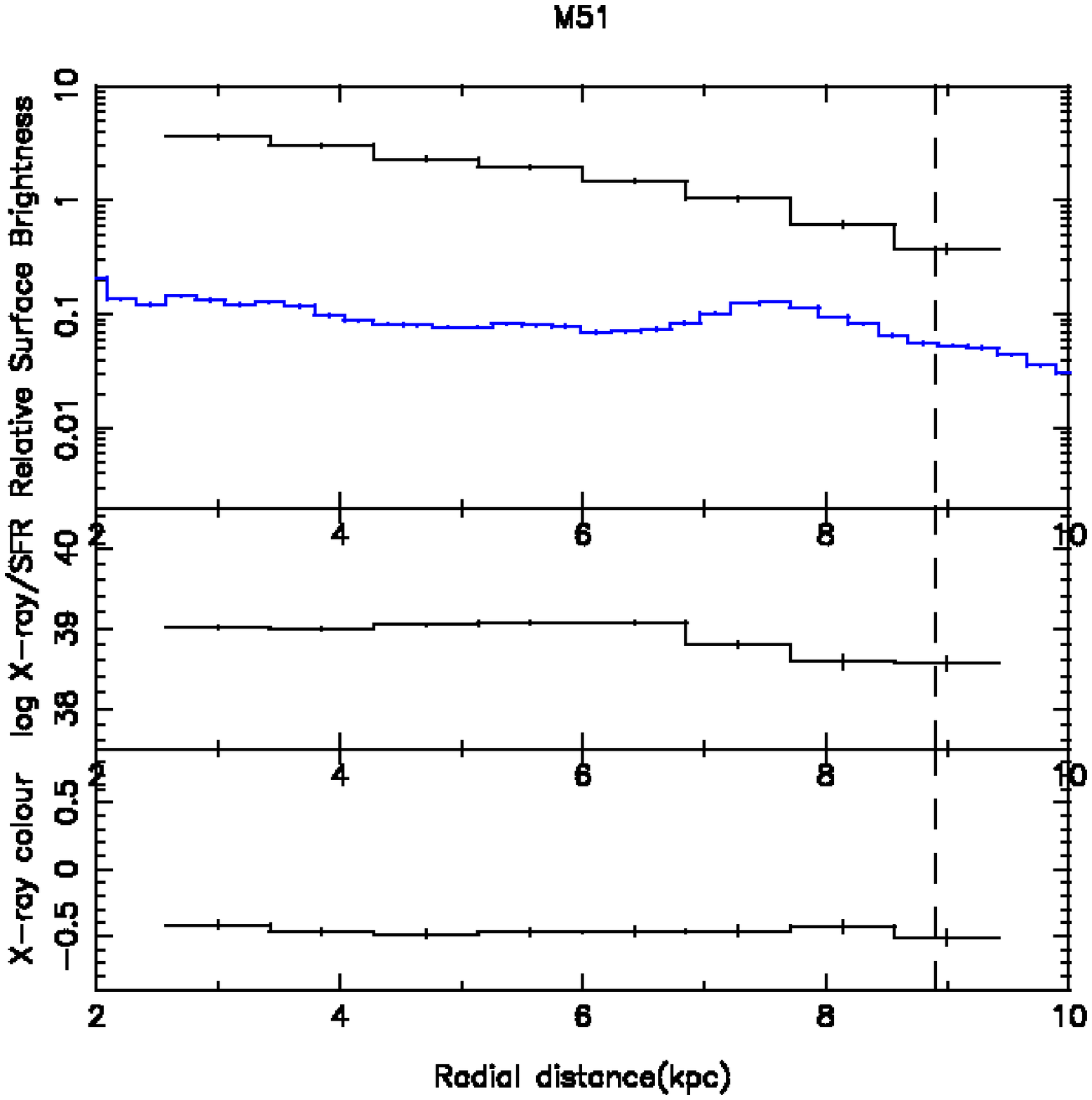}}}
\rotatebox{270}{\scalebox{0.4}{\includegraphics{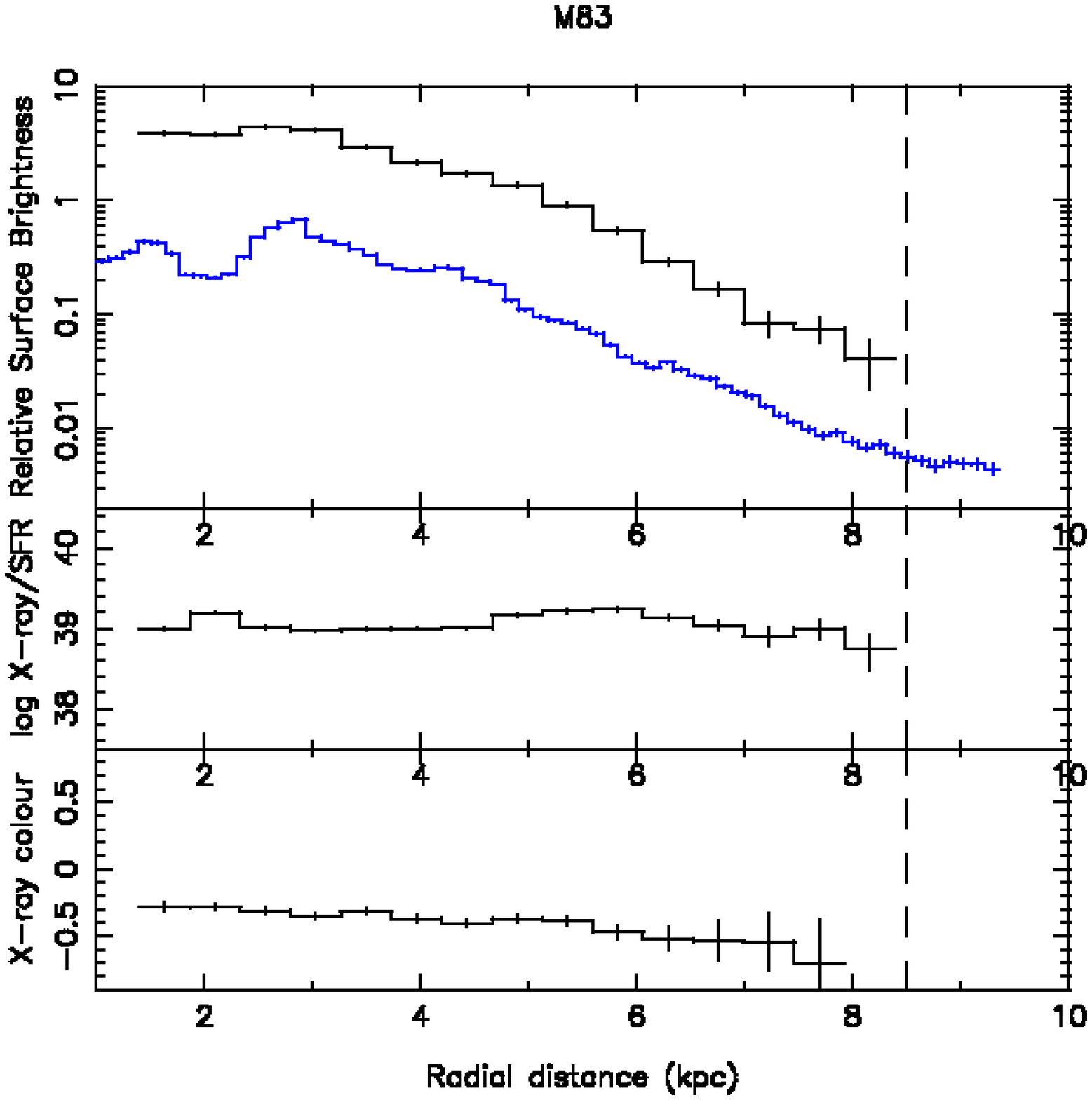}}}
\rotatebox{270}{\scalebox{0.4}{\includegraphics{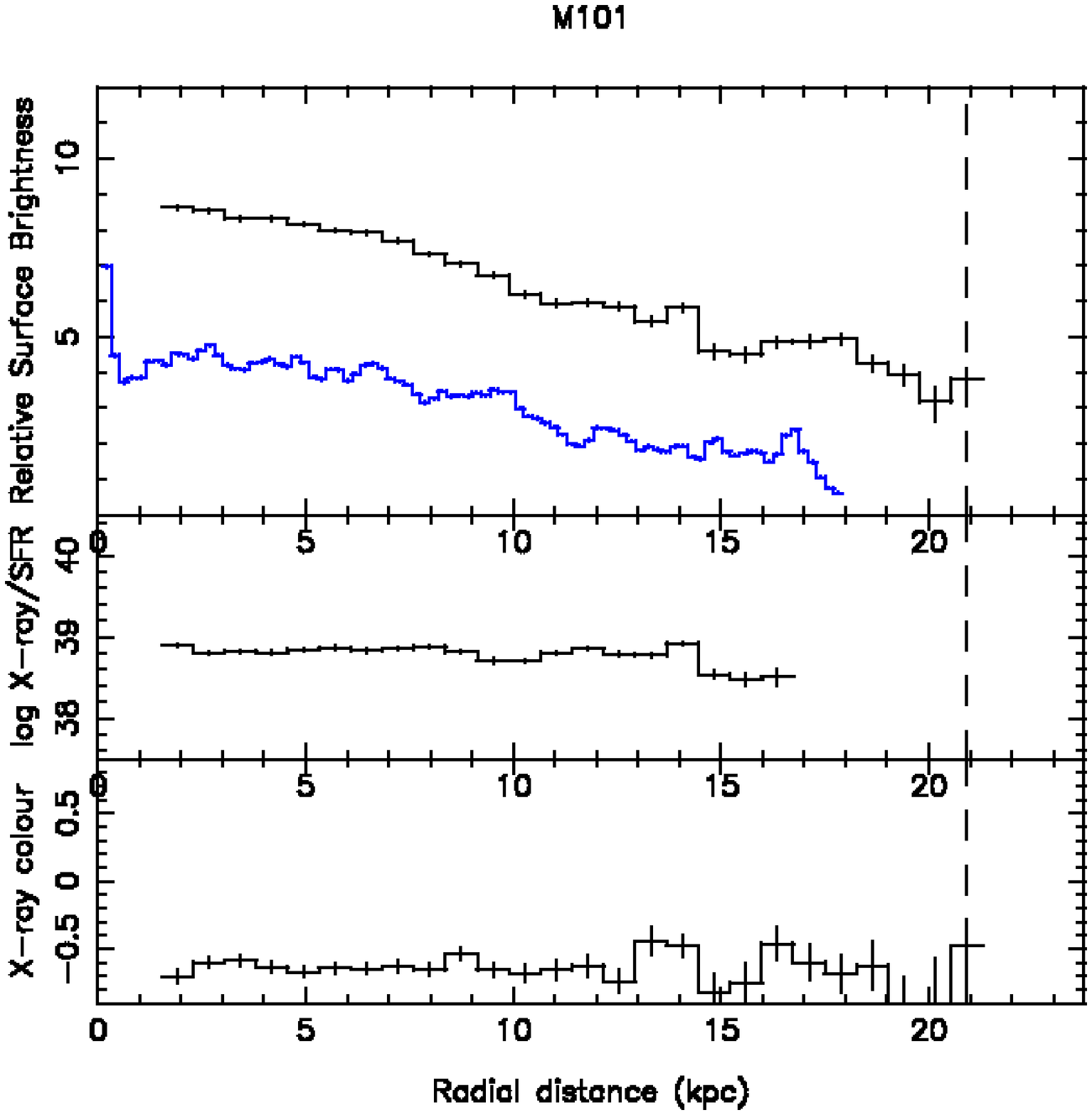}}}
\caption{A comparison of various radial dependencies for each
of the six galaxies in the full sample. In these plots the x-axis refers
to the major-axis radius scaled to kpc using the assumed distances
quoted in Table \ref{table:gal:details}. The vertical dashed line represents the 
extent of the X-ray extraction region scaled to kpc. For each galaxy, the following 
information is provided: {\it Top panel:}  The soft X-ray surface brightness 
versus radius (upper curve). The radial profile of the FUV emission after correction 
for internal extinction - taken from \citet{munoz07} (lower curve). 
{\it Middle panel:}  The ratio of the soft X-ray luminosity in $\ergsec$~pc$^{-2}$
(0.3-1 keV) to the local SFR in units of \Msun~yr$^{-1}$~pc$^{-2}$.  
{\it Bottom panel:} Variation in X-ray spectral hardness, (H-S)/(H+S), 
versus radius, where H refers to the 0.8--1.2 keV band and S to the 0.3--0.8 keV band.   
}
\label{fig:radial_plots}
\end{figure*}

Using the masked pn+MOS soft band images (such as in Fig. \ref{fig:im_big}), we 
have derived the radial profiles of the soft X-ray emission for all six galaxies 
in our sample.  The results are shown in Fig. \ref{fig:radial_plots} (upper panels). 
These X-ray radial profiles  are based on the signal extracted from 
{\it circular} annuli centred  on the galactic nucleus, with suitable 
normalization to allow for the regions excluded by the spatial masking.  
Since we wish to compare the X-ray radial distributions with published FUV 
and star-formation profiles obtained using {\it elliptical} annuli (matched
to the modest inclinations of these galaxies), we have stretched the scale of the
X-ray radial axis by a factor of typically 5--10\%, so as to account for the 
(marginally) different radial gradients obtained if one uses circular as opposed to 
slightly elliptical annuli. 


Figure \ref{fig:radial_plots} also shows the corresponding FUV radial brightness 
distributions taken from the tabulations in \citet{munoz07} (Table 2 in their
paper) except in the case of NGC 3184 where, since {\it GALEX} FUV data
are not available, we use \xmmn Optical Monitor UVW1 ($\approx 2680$ \AA )
measurements (without any internal extinction correction). 
These {\it GALEX} FUV data have been corrected for both foreground and internal 
extinction as discussed in \citet{munoz07}.

Visual comparison of the soft X-ray and FUV profiles suggest that radial fall-off
seen in the soft X-ray regime is either very comparable to that seen in the FUV
(\eg NGC 3184 and M83) or somewhat steeper (\eg NGC 300, M74, M51 and M101).
We further use the results of \citet{munoz07} to compare our current soft X-ray
measurements (after applying appropriate scaling factors to convert
the X-ray measurements into luminosity per unit disk area) with the inferred 
radial dependence of the SFR density (\ie SFR per unit disk area).  The results 
are also shown in Fig. \ref{fig:radial_plots} (middle panels).
We find that in the inner disks of our sample of late-type galaxies, the soft X-ray 
luminosity generated per unit SFR (hereafter the X-ray/SFR ratio) ranges from  
$2.5 \times10^{38}\ergsec$ (\Msun~yr$^{-1}$)$^{-1}$ 
in the case of NGC 300 up to $4 \times10^{38}\ergsec$ (\Msun~yr$^{-1}$) for M74
and finally up to $\sim 10^{39}\ergsec$ (\Msun~yr$^{-1}$)$^{-1}$ for M51, M83 and M101.
Note that these measurements refer to the soft-band 0.3--1.0 keV luminosity
{\it after excluding} the contribution of the most luminous point sources.
There is also the suggestion that the X-ray/SFR ratio
declines with galactocentric radius, at least in the case of NGC 300
and M51.

Whereas some correction has been applied to the FUV data to allow for the extinction 
intrinsic to the host galaxy, we have not applied a similar correction to the 
soft X-ray data.
Earlier we established that based on the spectral fitting, there was no clear evidence for 
any soft X-ray absorption intrinsic to the host galaxies. Any intrinsic 
soft X-ray extinction would be most pronounced in the inner galaxy regions potentially 
giving rise to a central down-turn in the soft X-ray radial distributions - but again 
there is no compelling evidence for such an 
effect\footnote{We can use the FUV extinction estimates
employed by \citet{munoz07} to predict the soft X-ray absorption using the scaling
A(FUV)/$N_{H}$ = $1.6 \times 10^{-21}$~mag~cm$^{-2}$, which assumes a Galactic
dust-to-gas ratio and extinction law with $R_{V}=3.1$ (\citealt{bohlin78}; 
\citealt{cardelli89}). This
leads to estimates of the intrinsic X-ray absorption in the inner disk regions in
the range $N_{H} = 5 - 12 \times 10^{20}$~cm$^{-2}$, across the set of galaxies.
Since these values are too high to be consistent with spectral fitting we can 
conclude either that the assumptions underlying the FUV to X-ray extinction 
scaling are not valid or, as seems likely, the FUV sources are more strongly 
embedded in obscuring material than 
the soft X-ray emission.}.  
As a final check we have 
looked for spectral trends versus radius by repeating the imaging analysis 
for two soft X-ray sub-bands, namely 0.3--0.8 keV and 0.8--1.2 keV. We 
then determine a spectral hardness ratio as (H-S)/(H+S), where H refers to 
the 0.8--1.2 keV band and S to the 0.3--0.8 keV band.  
Figure \ref{fig:radial_plots} (lower panels) shows the measured variations 
in spectral hardness versus radius. These curves show no marked spectral
variations towards the inner galaxy regions, thus confirming that intrinsic 
absorption does not strongly influence the soft X-ray measurements. 

The demarkation of the above sub-bands at 0.8 keV corresponds roughly 
to the position in the M51 and M83 spectra at which the cool and hot
thermal components contribute equally. It follows that the spectral hardness profiles 
in Fig. \ref{fig:radial_plots} track the ratio of these two thermal components 
as a function of galactocentric radius.  If we consider the three sources with 
best signal-to-noise,  we find no variation in spectral hardness versus radius 
in M101, a hint of spectral softening versus radius in M83 and the hint 
of an opposite trend in M51.


\begin{table*}
\caption{Derived properties of the galactic disks.}
\centering
\begin{tabular}{lcccccc}
\hline

Galaxy  & \multicolumn{2}{c} {Disk Radial Extent}  &  SFR   &  \lx (0.3--2 keV)$^{a}$ & Cool:Hot  \\
        & Inner (kpc)  & Outer (kpc)        &  $\Msun$ yr$^{-1}$  & $10^{39}\ergsec$ &  Flux Ratio \\
\hline
NGC300  &  0  &  2.9  &  0.19  &  0.11 &  -  \\
\\
M74  &  0  &  16.7  &  5.2  &  1.2  & $>5.4$ \\
\\
NGC3184  & 0 &  12.2  &  -  & 3.4  & $>4.1$ \\
\\
M51 &  2.4  &  8.9  &  7.5  &  6.8 & 1.3 \\
\\
M83 &  1.3  &  8.5  &  3.5  &  3.4 & 1.5\\
\\
M101 &  0  &  20.9  &  8.1  &  4.3 & 3.8 \\
\\
\hline    
\end{tabular}
\\
$^{a}$ - Luminosity after excluding sources with \lx $> 10^{37}\ergsec$ (0.3--6 keV)\\
\label{table:sfrdata}
\end{table*}


\begin{figure*}
\centering
\rotatebox{270}{\scalebox{0.5}{\includegraphics{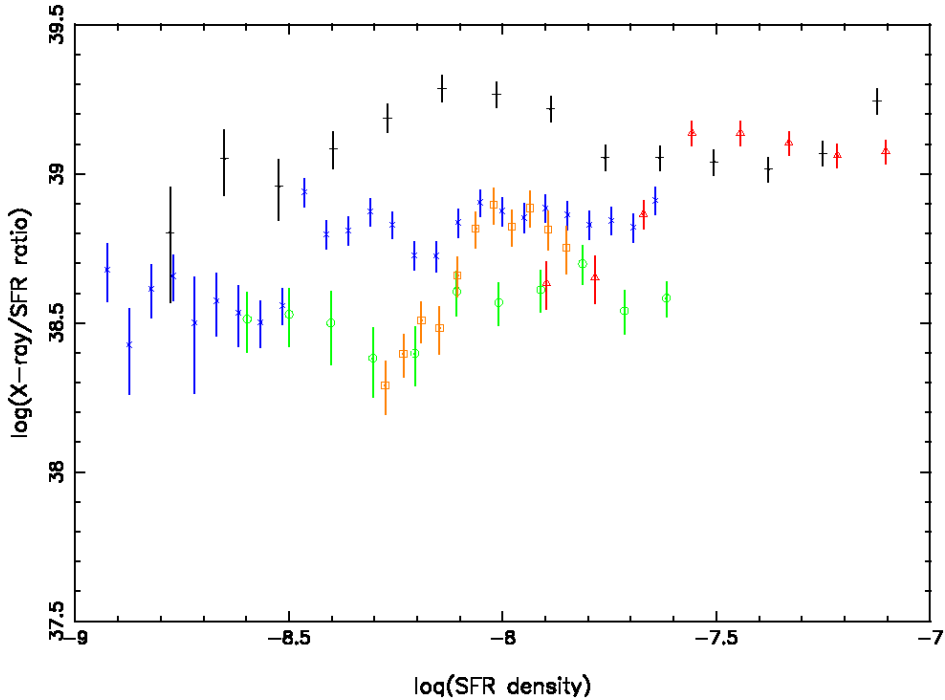}}}
\caption{The ratio of the soft X-ray luminosity in units of $\ergsec$~pc$^{-2}$ 
(0.3--2 keV) to the local SFR density in units of \Msun~yr$^{-1}$~pc$^{-2}$ 
versus the local SFR density.
The data points correspond to NGC300 (orange squares), M74 (green circles), 
M51 (red triangles), M83 (black points) and M101 (blue crosses).
}
\label{fig:sfrplot}
\end{figure*}


\section{Discussion}
\label{sec:disc}

A major objective of our investigation is to intercompare the X-ray properties and other 
characteristics of the extended disks of the six late-type galaxies which comprise our 
sample. To that end we have summarised some of the results from the current
analysis in Table \ref{table:sfrdata}.  In brief this table provides the following
information. We first define the extent of the disk region considered in X-rays. 
For this intercomparison we have chosen to exclude a central 1\arcm radius region
in both M51 and M83, so to avoid the complications associated with the AGN and/or
the nuclear starburst.  We have next calculated the total SFR within the specified 
disk region
by integrating under the exponential fits to the starburst density radial profiles
reported by \citet{munoz07}.  We next report the luminosity associated with the
disk by taking our soft band residual luminosity estimates and then correcting both 
to a common source exclusion threshold of  \lx $> 10^{37}\ergsec$ (0.3--6 keV)
(using essentially the same methodology as applied in \S\ref{sec:morphology})
and to a broader 0.3--2 keV band-pass (using the spectral models for the
residual emission regions discussed earlier\footnote{In the case of NGC300 we assume
a single component (0.2-keV) MEKAL spectrum and for M101 we use the spectral
model reported in W07.}).  These corrections were generally less than 
$20\%$  except for NGC300, where the luminosity increased by a factor of 2.2 as a result
of a significant additional contribution from discrete sources. Finally, we also note
the relative strengths of the cool (0.2-keV) and hot (0.65-keV) thermal components in 
terms of the ratio of their fluxes in the 0.3--2 keV (band), as inferred from the spectral
modelling (\S\ref{sec:spec:res}).

The results summarised in  Table \ref{table:sfrdata} establish that averaged over 
the defined disk regions, the X-ray/SFR ratio varies from a 
maximum of $\approx10^{39}\ergsec$ 
(\Msun~yr$^{-1}$)$^{-1}$ in the
case of M51 and M83 down to $\approx2 \times 10^{38}\ergsec$ (\Msun~yr$^{-1}$)$^{-1}$
in M74. However, these disk-average values mask a trend, which 
is evident in Fig. \ref{fig:sfrplot}, where we have plotted the X-ray/SFR ratio versus the
local SFR density. 
Above a local SFR density of $\sim3 \times 10^{-8}$ \Msun~yr$^{-1}$~pc$^{-2}$
(corresponding to the inner disks of M51 and M83), the X-ray/SFR ratio is consistent 
with an upper bound value of  $1-1.5 \times 10^{39}\ergsec$ (\Msun~yr$^{-1}$)$^{-1}$.
However, below a local SFR density of $\sim3 \times 10^{-8}$ \Msun~yr$^{-1}$~pc$^{-2}$,
there is significant scatter in the ratio, spanning a range up to a factor 5 below 
the maximum.  On the one hand some of this scatter appears to be due to the fact that
all five galaxies show  evidence for a fall-off in the X-ray/SFR ratio in their outer 
disks. On the other hand systematic errors in both the X-ray and FUV measurements 
will grow in 
importance as one tracks a signal of rapidly decreasing surface brightness against an 
essentially constant background. Nevertheless, these results do suggest that the efficiency
of X-ray production may be sensitive to the local level of star-formation activity
in the disk. 

In a recent study, \citet{mas08} (hereafter MH08) have investigated the temporal 
evolution of both
the soft X-ray and far infrared (FIR) luminosity following a burst of star formation.  
The starting point of their analysis is the adoption of a stellar population synthesis 
model 
which follows the evolution of a cluster of massive stars formed either at the same time 
(which MH08 refer to as an instanteous burst, IB) or at a constant rate over a period of
several tens of Myr (an extended burst, EB).
In either scenario, diffuse soft X-ray emission is produced through the heating of bubbles
within the interstellar medium to temperatures of $\sim 10^{6-7}$ K  as a result of 
the mechanical energy input from the winds of massive stars created in the starburst
or the eventual destruction of such stars in supernovae. Their model also includes 
the soft X-rays   
emitted by individual SNR during the adiabatic expansion phase of their evolution, but
excludes any contribution from the stellar atmospheres (which in any case would
be extremely small) or accreting high-mass X-ray binaries created in the region.
MH08 (see their Figure 1) find that the soft X-ray luminosity increases rapidly 
during the first few Myr following the onset of starburst activity, largely through
the input power of the stellar winds of the most massive stars. After the first 3 Myr,
the most massive stars produce supernovae and thereafter it is the mechanical
energy of such explosions which dominates the energy budget. In the IB models the
soft X-ray luminosity peaks at $\sim 3$ Myr, whereas in the EB models it continues to 
rise, 
albeit more gradually, until eventually an equilibrium is established between the 
formation and destruction of stars (after typically $\sim40$ Myr for solar metallicities).

In comparing our present measurements with the predictions of MH08
it is important to bear in mind that we have used SFR estimates based on FUV
measurements (\citealt{munoz07}), which in turn rely on the UV-to-SFR calibration 
suggested by \citet{kennicutt98}. As noted by MH08, the
\citet{kennicutt98} calibration assumes star formation over a wide 0.1--100 
\Msun~mass range, whereas the SFR values they quote are for a more
restricted 2--120 \Msun~range (which has a more direct bearing on the resulting
soft X-ray and FIR luminosity). Assuming a Salpeter initial mass function, one
obtains the scaling  SFR(2-120\Msun)/ SFR(0.1-100\Msun) $= 0.293$, implying
that our estimates of the X-ray/SFR ratio should be increased by a factor 3.4
when making a comparison with the MH08 predictions.
Our upper-bound estimate, after applying the above scaling, becomes 
$\sim 5 \times 10^{39}\ergsec$ 
(\Msun~yr$^{-1}$)$^{-1}$, which correponds to the X-ray/SFR ratio predicted
some 10 Myr after the onset of an extended burst of star-formation, assuming
a $1\%$ efficiency in the conversion of mechanical energy into
X-ray emission.  Although it can be argued that star formation
in the disks of late-type spiral galaxies generally proceeds over long-periods
of time, the underlying activity pattern is presumably one of localised
bursts of star formation, so a match with ``young'' extended bursts is perhaps
not unreasonable. Of course, in matching the observations to the predictions,
the efficiency of X-ray production and the duration of the EB episode are inversely 
linked, for example the above X-ray/SFR ratio is also reached after 30 Myr in 
the EB scenario, 
provided the energy conversion efficiency is scaled down to just $0.2\%$. 
Clearly a more detailed study, taking into account additional information
relating to the current star formation rate and the star formation history 
relevant to each galaxy disk, might well pin down
these parameters - but such an analysis is beyond the scope of the current paper.  

A number of recent studies have investigated whether the integrated soft 
(nominally 0.5-2 keV) 
or hard (2-10 keV) X-ray luminosities measured in galaxies not dominated by 
a central AGN  might serve as a proxy for the underlying galaxy-wide SFR
(\eg \citealt{grimm03}; \citealt{ranalli03}; \citealt{gilfanov04}), which might 
then be applicable to more distant objects (\eg 
\citealt{rosa07}). If we consider the \citet{ranalli03}
calibration of the SFR in terms of the integrated soft X-ray luminosity
(their equation 14), apply the corrections noted by MH08 and also scale 
the X-ray luminosity from a 0.5--2 keV bandpass to the 0.3--2 keV bandpass
(which increases the luminosity by a factor of $1.1-1.5$ depending on the spectral
assumption), we obtain an X-ray/SFR ratio in the range $10-14 \times 10^{39}\ergsec$ 
(\Msun~yr$^{-1}$)$^{-1}$ which is factor $2-2.8$ higher than the
(upper-bound) value derived from the current analysis. The difference
is, of course, due to the fact that \citet{ranalli03} use the integrated
soft X-ray luminosity of the entire galaxy, which includes a very substantial
contribution from the bright discrete source population.  In fact
the ratio of the total galaxy luminosity to the residual galaxy
luminosity over the 0.3-2 keV band for our current set of galaxies
(see Table \ref {table:gal_xray}) is typically in the range $2-2.5$, 
consistent with this picture.  It follows that when using the \citet{ranalli03}
calibration to convert soft X-ray measurements to SFR rate estimates,
an implicit assumption is that {\it both} the rate of production of X-ray binaries 
following a starburst and the subsequent balance between point source and 
diffuse luminosity is comparable to that pertaining in local galaxies.
 
As discussed above there is evidence that the X-ray/SFR ratio falls as the local SFR
density declines, implying that the interstellar environment influences either the
starburst and/or the efficiency with which its energy is reprocessed into the X-rays.  
Factors which may potentially be relevant include the local gas density (relevant to the 
conversion of mechanical energy into heat via shocks) and the local metallicity 
(which may, for example, influence the evolution timescales of massive stars, 
their stellar wind properties
and the emissivity of the X-ray plasma heated by the starburst). It is also notable that 
the galaxies in our current sample with the highest X-ray/SFR ratios are those
with the strongest weighting towards
hot plasma (as measured by the cool:hot flux ratios in Table \ref{table:sfrdata}). 
In this context it would seem entirely plausible that X-ray production efficiency 
and ``mean'' plasma temperature are coupled quantities.   Although the current data 
are too restricted to merit further consideration of these issues, such effects
may well be relevant when attempting to fine tune the calibration
of the soft X-ray luminosity versus SFR relation for specific
applications in which the bright source component is spatially resolvable.

Star formation in the disks of spiral galaxies is thought to be triggered by the passage 
of a spiral density wave through the ISM. The resultant massive star formation results 
in copious FUV emission which in turn is rapidly reprocessed into \Ha~and FIR
emission. As discussed above, the mechanical energy input from stellar winds and 
supernovae 
eventually gives rise to X-ray emission.  This would seem to give an immediate explanation
of the strong correlation observed between the infrared, optical and UV star-formation
tracers and the soft X-ray emission seen in spiral galaxies ({\it e.g.,} \citealt{read01};
\citealt{kuntz03}; \citealt{tyler04}, W07). However, the underlying timescales are not 
identical. UV and \Ha~emission will be produced on a timescale commensurate
with the lifetime of the most massive stars ($\sim 3 \times 10^{6}$ years), whilst 
according to model predictions (\citealt{leitherer95}; \citealt{cervino02};
MH08), it takes approximately ten times longer to maximise the diffuse X-ray emission.  
Given the delay between the peak in UV/H${\alpha}$ 
emission and the X-ray heating, one might predict a spatial offset between the diffuse 
X-ray emission and the other spiral tracers more closely tied to the passage
of the spiral density wave. For example, W07 argue that in the case of M101
one might predict a rotational lag of $\approx 24^{\circ}$ at a galactocentric
radius of $r\approx7.5$ kpc (given a rotational velocity $v_{rot}\approx200~\rm~km~s^{-1}$
and assuming that the spiral pattern corotates with the disk material at $r\approx15$ kpc).
However, lags of this magnitude are not generally evident in the X-ray morphology of 
late-type galaxies (\citealt{tyler04}).  

Assuming a gas filling factor $\eta \sim$ 1 and 
a spatial extent perpendicular to the plane of the galaxy of 0.5 kpc,  we earlier derived 
radiative cooling timescales in the range $2-10\times 10^{8}$yr  (Table 
\ref{table:gas_physics}). These estimates are comparable to the galaxy rotational 
periods in
the region sampled by the current X-ray measurements and are clearly inconsistent with 
the presence of narrow X-ray spiral arm features. Matching the radiative timescales
to the ``young'' extended burst scenario discussed earlier would require $\eta<< $1,
consistent with a clumpy thin-disk distribution for the X-ray emitting plasma most 
closely associated
with the spiral arms. In practice, however, the observed X-ray emission will
encompass a complex web of diffuse features extending over a range of spatial scales
including superbubbles and, in the most active regions, outflows into the 
lower galactic halo (\eg \citealt{strickland04a}; \citealt{strickland04b}). 
Where there is no confinement by chimneys or similar structures in the 
ISM, energy losses arising from adiabatic expansion of the hot gas in the disk may 
also help localise the spiral arm component (W07). Relatively luminous discrete sources 
including young SNRs, some perhaps linked to disk population stars in interarm regions, 
together with aggregations of less luminous objects, add further to the mix. 


\section{Conclusions}
\label{sec:conc}

We have used \xmmn observations in a study of the extended extranuclear disks 
of six nearby face-on spiral galaxies.  Using a novel spatially 
masking technique to minimise the impact of the brightest discrete sources, 
we have investigated both the spatial morphology and the spectral properties 
of the residual soft X-ray emission emanating from the disk regions.
The strong correlation found between soft X-ray and FUV images in terms 
of the tracing of specific spiral features and the overall radial extent of the emission
unambigously establishes a close link between the soft X-ray emission and 
recent star formation. 

More detailed comparison of the radial profiles
of soft X-ray and FUV surface brightness distributions establishes
an X-ray/SFR ratio of $\sim5 \times10^{39}\ergsec$ (\Msun~yr$^{-1}$)$^{-1}$ 
(referred to the 0.3--2 keV band and where the SFR applies to stars in the 
mass range 2--120 \Msun) for the inner disks of M51 and M83.  
This is roughly a factor 2.5 below the soft X-ray to SFR calibration report 
by \citet{ranalli03}, consistent with the fact that our estimate excludes 
the contribution of discrete sources with \lx $> 10^{37}\ergsec$ (0.3--6 keV).  
Our measured X-ray/SFR ratio matches that predicted some 10 
Myr after the onset of an extended burst of star formation in the
models of MHO7, when the efficiency of the conversion of mechanical
energy input to X-rays is set at $\sim 1\%$.  Our observations
suggest a fall-off in the X-ray/SFR ratio as the local SFR
density declines, implying that the interstellar environment influences either the
starburst and/or the efficiency with which its energy is reprocessed into the X-rays. 

With the bright sources excluded, the spectra of the residual galaxy regions
are, in the main, well matched by a two-temperature thermal plasma model. 
The characteristic temperatures of $\approx0.2$ and $\approx0.65$ keV are in line with 
published results for other spiral and starburst galaxies. 
The relative strengths of these two components varies across the sample, with
the galaxies having the highest X-ray/SFR ratio characterised by a higher ``mean'' 
temperature.  The physical properties of the gas found in the disk are shown to be 
consistent with a clumpy thin-disk distribution presumably composed of diffuse structures
such as superbubbles together with individual SNRs and other source aggregations.  

Future observations should allow more detailed investigation of a wide range of factors,
for example the local gas density and the local metallicity, which might influence 
the X-ray/SFR ratio measured in late-type galaxy disks.


\section*{Acknowledgments}

We thank the referee for providing very useful suggestions for
the improvement of this paper. The X-ray data presented in this
paper were obtained from the public archive maintained by the \xmmn
science operations centre at ESAC. The {\sl GALEX} data were
obtained from the Multimission Archive at the Space Telescope 
Science Institute (MAST). RAO acknowledges the receipt of a 
PPARC/STFC research studentship.

\label{lastpage}

{}

\end{document}